\documentclass[twocolumn,a4paper,allowtoday,accepted=2021-07-11]{quantumarticle}

\pdfoutput=1

\usepackage{dcolumn}
\usepackage{bm}
\usepackage{qcircuit}
\usepackage{graphicx}
\usepackage{physics}
\usepackage{mathtools}
\usepackage{tikz}
\usepackage{circuitikz}
\usepackage[numbers,sort&compress]{natbib}
\usepackage{url}
\usepackage{hyperref} %
\usepackage{tikzit}
\usepackage{float}
\usepackage{cleveref}
\usepackage[caption=false]{subfig}
\usepackage{amsfonts}

\tikzstyle{normal}=[line width=0.9pt]
\tikzstyle{doubled}=[line width=1.5pt] %
\tikzstyle{boldedge}=[doubled]

\tikzstyle{dot}=[inner sep=0mm,minimum width=2mm,minimum height=2mm,draw,shape=circle]  

\tikzstyle{black dot}=[dot,fill=black]
\tikzstyle{white dot}=[dot,fill=white]

\tikzstyle{discarding}=[fill=white, draw=black, shape=circle, style=upground]
\tikzstyle{smalldiscarding}=[fill=white, draw=black, shape=circle, style=upground, scale=0.5]
\tikzstyle{backdiscard}=[fill=white, draw=black, shape=circle, style=downground]
\tikzstyle{smallbackdiscard}=[fill=white, draw=black, shape=circle, style=downground, scale=0.5]
\tikzstyle{state}=[fill=white, draw=black, style=point, tikzit shape=rectangle]
\tikzstyle{kstate}=[fill=white, draw=black, style=kpoint, tikzit shape=rectangle]
\tikzstyle{kstateconj}=[fill=white, draw=black, style=kpoint conjugate, tikzit shape=rectangle]
\tikzstyle{kstateBIG}=[fill=white, draw=black, style=big kpoint, tikzit shape=rectangle]
\tikzstyle{effect}=[fill=white, draw=black, style=copoint]
\tikzstyle{keffect}=[fill=white, draw=black, style=kpoint adjoint]
\tikzstyle{keffectconj}=[fill=white, draw=black, style=kpoint transpose]
\tikzstyle{morphdag}=[style=map]
\tikzstyle{morph}=[style=mapdag]
\tikzstyle{morphtrans}=[style=mapconj]
\tikzstyle{morphconj}=[style=maptrans]
\tikzstyle{CPMmorph}=[style=dmapdag]
\tikzstyle{CPMmorphconj}=[style=dmaptrans]
\tikzstyle{CPMkstate}=[fill=white, draw=black, style=kpoint, tikzit shape=rectangle, doubled]
\tikzstyle{CPMkstateconj}=[fill=white, draw=black, style=kpoint conjugate, tikzit shape=rectangle, doubled]
\tikzstyle{CPMkstateBIG}=[fill=white, draw=black, style=big kpoint, tikzit shape=rectangle, doubled]
\tikzstyle{CPMkeffect}=[fill=white, draw=black, style=kpoint adjoint, doubled]
\tikzstyle{CPMkeffectconj}=[fill=white, draw=black, style=kpoint transpose, doubled]
\tikzstyle{bigmorph}=[style=medium map dag]
\tikzstyle{leak}=[style=tinypoint, regular polygon rotate = -90]
\tikzstyle{leakfill}=[style=tinypoint, regular polygon rotate = -90,fill=black]
\tikzstyle{Z} = [style=green dot, fill=green]
\tikzstyle{X} = [style=red dot, fill=red]
\tikzstyle{black_dot} = [style=black dot]
\tikzstyle{white_dot} = [style=white dot]
\tikzstyle{qwhite_dot} = [style=white ddot]
\tikzstyle{qblack_dot} = [style=black ddot]
\tikzstyle{qgrey_dot} = [style=gray ddot]
\tikzstyle{Zphase} = [style=phase dimensions, fill=green]
\tikzstyle{whitephase} = [style=white phase dot]
\tikzstyle{qredphase} = [style=white phase ddot, fill=red]
\tikzstyle{qgreenphase} = [style=white phase ddot, fill=green]
\tikzstyle{had} = [style=hadamard, doubled]
\tikzstyle{classhad} = [style=hadamard]

\tikzstyle{dottededge}=[-, dotted]
\tikzstyle{double edge}=[-, style=boldedge, draw=black, tikzit draw={rgb,255: red,191; green,0; blue,64}]

\usetikzlibrary{calc, positioning, shapes.geometric}
\usetikzlibrary{
	arrows,
	shapes,
	decorations,
	intersections,
	backgrounds,
	positioning,
	circuits.ee.IEC
	}

\newcommand{\tikzfigscale}[2]{\scalebox{#1}{\input{#2.tikz}}}

\makeatletter
\renewcommand*\env@matrix[1][*\c@MaxMatrixCols c]{%
  \hskip -\arraycolsep
  \let\@ifnextchar\new@ifnextchar
  \array{#1}}
\makeatother

\begin{document}

\author{Oscar Higgott}
\email{oscar.higgott.18@ucl.ac.uk}
\affiliation{Department of Physics and Astronomy, University College London, Gower Street, London WC1E 6BT, United Kingdom}
\author{Matthew Wilson}
\affiliation{Department of Physics and Astronomy, University College London, Gower Street, London WC1E 6BT, United Kingdom}
\affiliation{Department of Computer Science, University of Oxford, Oxford OX1 3QD, United Kingdom}
\author{James Hefford}
\affiliation{Department of Physics and Astronomy, University College London, Gower Street, London WC1E 6BT, United Kingdom}
\affiliation{Department of Computer Science, University of Oxford, Oxford OX1 3QD, United Kingdom}
\author{James Dborin}
\affiliation{Department of Physics and Astronomy, University College London, Gower Street, London WC1E 6BT, United Kingdom}
\affiliation{London Centre for Nanotechnology, University College London,
Gordon St., London WC1H 0AH, United Kingdom}
\author{Farhan Hanif}
\affiliation{Department of Physics and Astronomy, University College London, Gower Street, London WC1E 6BT, United Kingdom}
\author{Simon Burton}
\affiliation{Department of Physics and Astronomy, University College London, Gower Street, London WC1E 6BT, United Kingdom}
\author{Dan E. Browne}
\affiliation{Department of Physics and Astronomy, University College London, Gower Street, London WC1E 6BT, United Kingdom}

\title{Optimal local unitary encoding circuits for the surface code}

\date{\today}

\begin{abstract}
    The surface code is a leading candidate quantum error correcting code, owing to its high threshold, and compatibility with existing experimental architectures. Bravyi \textit{et al.}~\cite{bravyi2006lieb} showed that encoding a state in the surface code using local unitary operations requires time at least linear in the lattice size $L$, however the most efficient known method for encoding an unknown state, introduced by Dennis \textit{et al.}~\cite{dennis2002topological}, has $O(L^2)$ time complexity. Here, we present an optimal local unitary encoding circuit for the planar surface code that uses exactly $2L$ time steps to encode an unknown state in a distance $L$ planar code. We further show how an $O(L)$ complexity local unitary encoder for the toric code can be found by enforcing locality in the $O(\log L)$-depth non-local renormalisation encoder. We relate these techniques by providing an $O(L)$ local unitary circuit to convert between a toric code and a planar code, and also provide optimal encoders for the rectangular, rotated and 3D surface codes. 
    Furthermore, we show how our encoding circuit for the planar code can be used to prepare fermionic states in the compact mapping, a recently introduced fermion to qubit mapping that has a stabiliser structure similar to that of the surface code and is particularly efficient for simulating the Fermi-Hubbard model. 
\end{abstract}

\maketitle

\section{\label{sec:intro}Introduction}

One of the most promising error correcting codes for achieving fault-tolerant quantum computing is the surface code, owing to its high threshold and low weight check operators that are local in two dimensions~\cite{kitaev2003fault, dennis2002topological}. The stabilisers of the surface code are defined on the faces and sites of a $L \times L$ square lattice embedded on either a torus (the \textit{toric} code) or a plane (the \textit{planar} code). The toric code encodes two logical qubits, while the planar code encodes a single logical qubit.

An important component of any quantum error correction (QEC) code is its encoding circuit, which maps an initial product state of $k$ qubits in arbitrary unknown states (along with $n-k$ ancillas) to the same state on $k$ logical qubits encoded in a quantum code with $n$ physical qubits. The encoding of logical states has been realised experimentally for the demonstration of small-scale QEC protocols using various codes~\cite{ chiaverini2004realization, lu2008experimental, schindler2011experimental, taminiau2014universal, waldherr2014quantum, nigg2014quantum, kelly2015state, ofek2016extending,  cramer2016repeated, linke2017fault, vuillot2017error, roffe2018protecting,gong2021experimental}, however one of the challenges of realising larger-scale experimental demonstrations of QEC protocols is the increasing complexity of the encoding circuits with larger system sizes, which has motivated the recent development of compiling techniques that reduce the number of noisy gates in unitary encoding circuits~\cite{xu2021variational}.

Encoding circuits can also be useful for implementing fermion-to-qubit mappings~\cite{seeley2012bravyi}, an important component of quantum simulation algorithms, since some mappings introduce stabilisers in order to mitigate errors~\cite{jiang2019majorana} or enforce locality in the transformed fermionic operators~\cite{bravyi2002fermionic,verstraete2005mapping,steudtner2019quantum,havlivcek2017operator}. Local unitary encoding circuits provide a method to initialise and switch between mappings without the need for ancilla-based stabiliser measurements and feedback.

The best known local unitary circuits for encoding an unknown state in the surface code are far from optimal. Bravyi \textit{et al.}~\cite{bravyi2006lieb} showed that any local unitary encoding circuit for the surface code must take time that is at least linear in the distance $L$, however the most efficient known local unitary circuit for encoding an unknown state in the surface code was introduced by Dennis \textit{et al.}~\cite{dennis2002topological}, and requires $\Omega(L^2)$ time to encode an unknown state in a distance $L$ planar code. Aguado and Vidal~\cite{aguado_MERA} introduced a Renormalisation Group (RG) unitary encoding circuit for preparing and unknown state in the toric code with $O(\log L)$ circuit depth, however their method requires non-local gates. More recently, Aharonov and Touati provided an $\Omega(\log L)$ lower bound on the circuit depth of preparing toric code states with non-local gates, demonstrating that the RG encoder is optimal in this setting~\cite{aharonov2018quantum}, and an alternative approach for preparing a specific state in the toric code with non-local gates and depth $O(\log L)$ was recently introduced in Ref.~\cite{liao2021quantum}. Dropping the requirement of unitarity, encoders have been found that use stabiliser measurements~\cite{lodyga2015simple, horsman2012surface,li2015magic} or local dissipative evolution~\cite{dengis2014optimal}, and it has been shown that local dissipative evolution cannot be used to beat the $\Omega(L)$ lower bound for local unitary encoders~\cite{konig2014generating}. If only the logical $\bar{\ket{0}}$ state is to be prepared, then stabiliser measurements~\cite{dennis2002topological} can be used, as well as optimal local unitaries that either use adiabatic evolution~\cite{hamma2008adiabatic} or a mapping from a cluster state~\cite{brown2011generating}. However, encoding circuits by definition should be capable of encoding an arbitrary unknown input state.

In this work, we present local unitary encoding circuits for both the planar and toric code that take time linear in the lattice size to encode an unknown state, achieving the $\Omega(L)$ lower bound given by Bravyi \textit{et al.}~\cite{bravyi2006lieb}. Furthermore, we provide encoding circuits for rectangular, rotated and 3D surface codes, as well as a circuit that encodes a toric code from a planar code. Our circuits also imply optimal encoders for the 2D color code~\cite{kubica2015unfolding}, some 2D subsystem codes~\cite{bombin2012universal, bravyi2012subsystem} and any 2D translationally invariant topological code~\cite{bombin2012universal}. On many Noisy Intermediate-Scale Quantum (NISQ)~\cite{preskill2018quantum} devices, which are often restricted to local unitary operations, our techniques therefore provide an optimal method for experimentally realising topological quantum order.
Another advantage of using a unitary encoding circuit is that it does not require the use of ancillas to measure stabilisers, therefore providing a more qubit efficient method of preparing topologically ordered states ($2\times$ fewer qubits are required to prepare a surface code state of a given lattice size).
Finally, we show how our unitary encoding circuits for the planar code can be used to construct $O(L)$ depth circuits to encode a Slater determinant state in the compact mapping~\cite{derby2021compact}, which can be used for the simulation of fermionic systems on quantum computers.

\section{Stabiliser codes}\label{sec:stab}

An $n$-qubit Pauli operator $P=\alpha P_n$ where $P_n\in\{I,X,Y,Z\}^{\otimes n}$ is an $n$-fold tensor product of single qubit Pauli operators with the coefficient $\alpha\in\{\pm 1, \pm i\}$. The set of all $n$-qubit Pauli operators forms the $n$-qubit Pauli group $\mathcal{P}_n$. The \textit{weight} $\mathrm{wt}(P)$ of a Pauli operator $P\in\mathcal{P}_n$ is the number of qubits on which it acts non-trivially. Any two Pauli operators commute if an even number of their tensor factors commute, and anti-commute otherwise.

Stabiliser codes~\cite{gottesman1997stabilizer} are defined in terms of a stabiliser group $\mathcal{S}$, which is an abelian subgroup of $\mathcal{P}_n$ that does not contain the element $-I$. Elements of a stabiliser group are called \textit{stabilisers}. Since every stabiliser group is abelian and Pauli operators have the eigenvalues~$\pm 1$, there is a joint $+1$-eigenspace of every stabiliser group, which defines the stabiliser code.

The \textit{check operators} of a stabiliser code are a set of generators of $\mathcal{S}$ and hence all measure $+1$ if the state is uncorrupted. Any check operator $M$ that anticommutes with an error $E$ will measure -1 (since $ME\ket{\psi}=-EM\ket{\psi}=-E\ket{\psi}$). The centraliser $C(\mathcal{S})$ of $\mathcal{S}$ in $\mathcal{P}_n$ is the set of Pauli operators which commute with every stabiliser. If an error $E\in C(\mathcal{S})$ occurs, it will be undetectable. If $E\in\mathcal{S}$, then it acts trivially on the codespace, and no correction is required. However if $E\in C(\mathcal{S})\setminus \mathcal{S}$, then an undetectable logical error has occurred. The distance $d$ of a stabiliser code is the smallest weight of any logical operator.

A stabiliser code is a Calderbank-Shor-Steane (CSS) code if there exists a generating set for the stabiliser group such that every generator is in $\{I,X\}^n\cup \{I,Z\}^n$.

\section{The Surface Code}

\begin{figure}
    \centering
    \subfloat[]{
    \includegraphics[width=0.58\columnwidth,trim={0 0 0 0}]{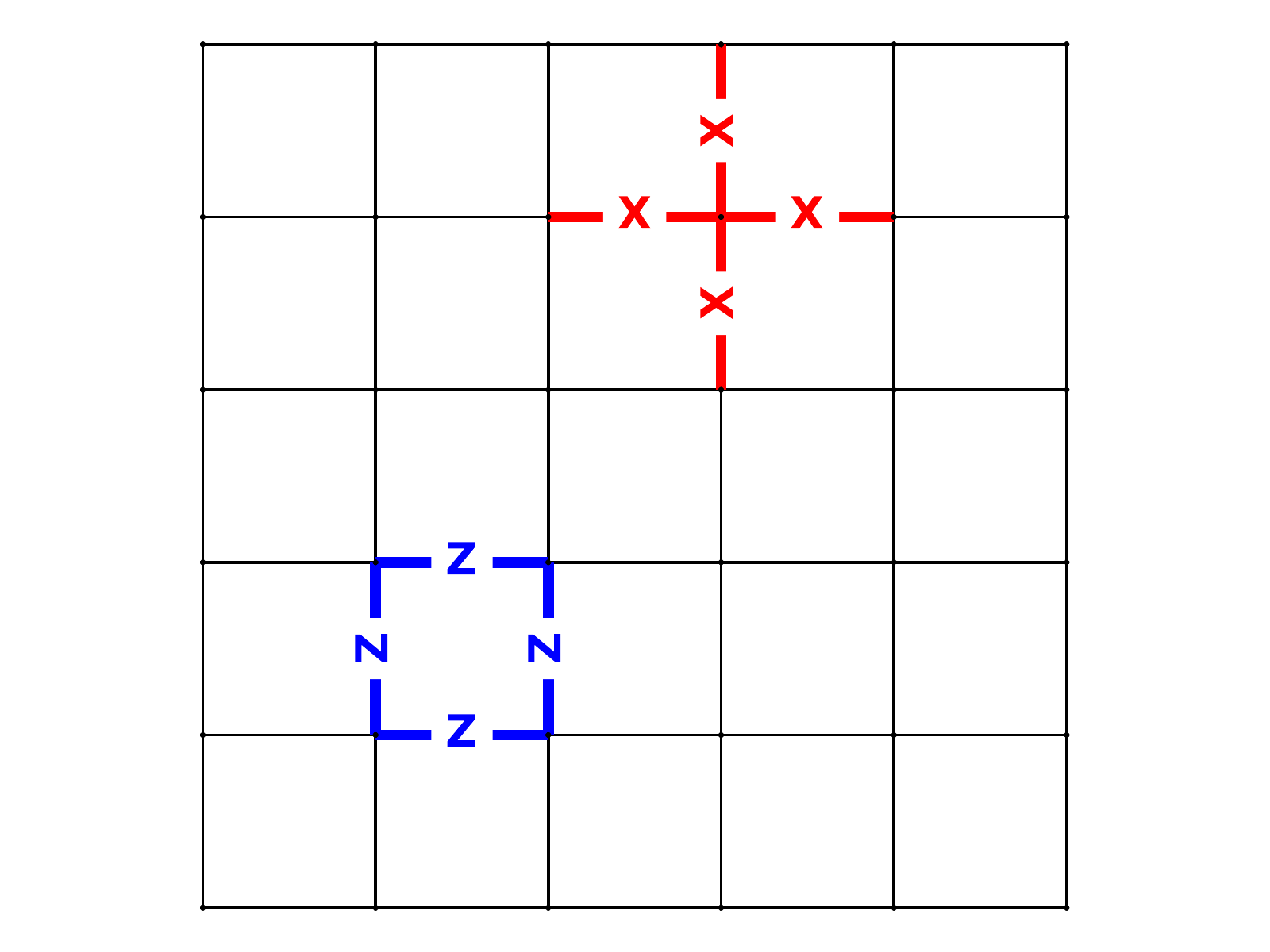}
    }
    \subfloat[]{
    \includegraphics[width=0.4\columnwidth,trim={2cm 0 4cm 0}]{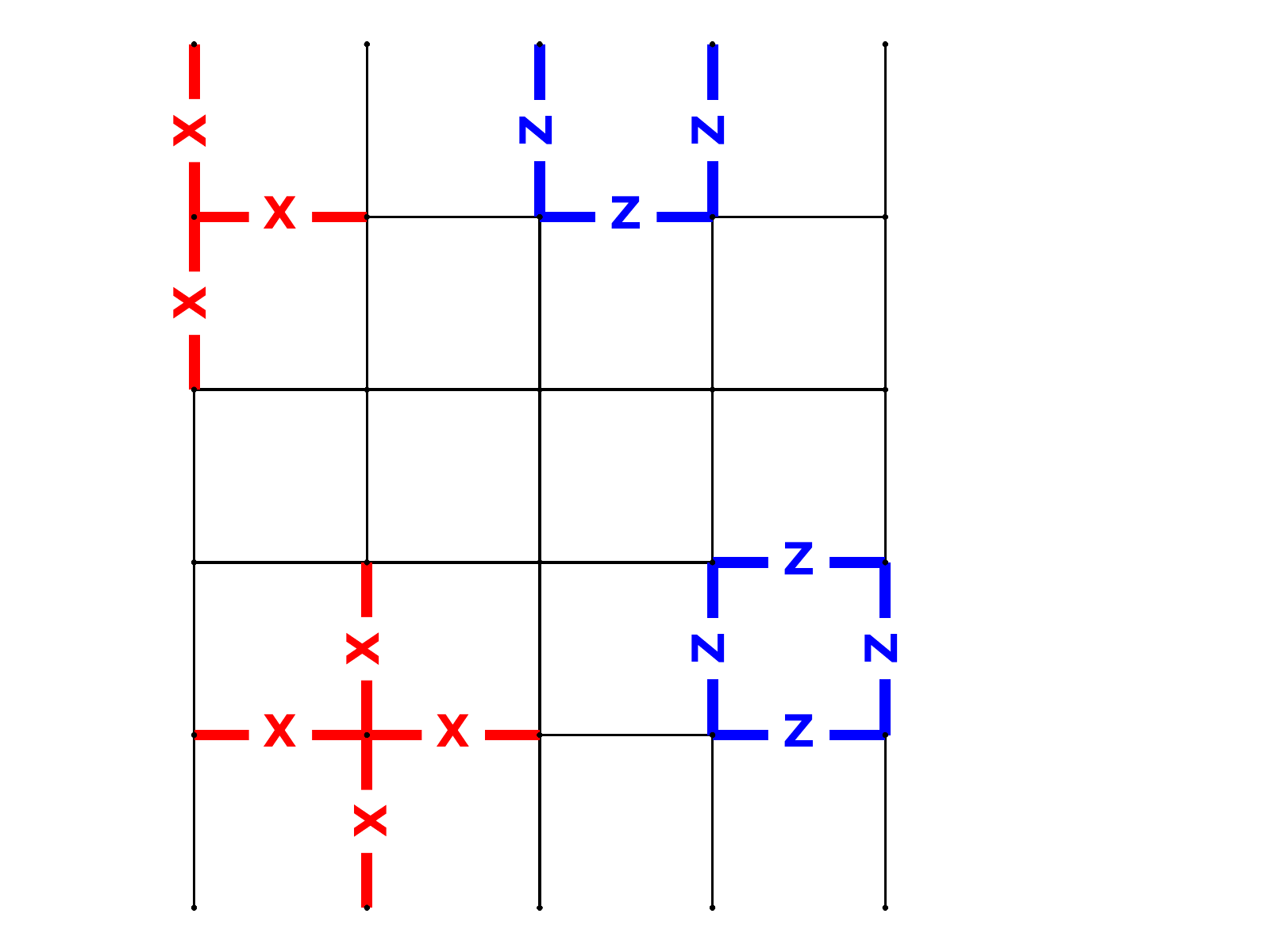}
    }
    \caption{The check operators for (a) the toric code and (b) the planar code. Opposite edges in (a) are identified and each edge corresponds to a qubit.}
    \label{fig:surface_code}
\end{figure}

The surface code is a CSS code introduced by Kitaev~\cite{kitaev2003fault, dennis2002topological}, which has check operators defined on a square lattice embedded in a two-dimensional surface. Each \textit{site} check operator is a Pauli operator in $\{I,X\}^n$ which only acts non-trivially on the edges adjacent to a vertex of the lattice. Each $\textit{plaquette}$ check operator is a Pauli operator in $\{I,Z\}^n$ which only acts non-trivially on the edges adjacent to a face of the lattice. In the toric code, the square lattice is embedded in a torus, whereas in the planar code the lattice is embedded in a plane, without periodic boundary conditions (see \Cref{fig:surface_code}). These site and plaquette operators together generate the stabiliser group of the code. While the toric code encodes two logical qubits, the surface code encodes a single logical qubit.

\section{Encoding an unknown state}\label{sec:encoding_unknown}

We are interested in finding a unitary encoding circuit that maps a product state $\ket{\phi_0}\otimes\ldots\otimes\ket{\phi_{k-1}}\otimes \ket{0}^{\otimes (n-k)}$ of $k$ physical qubits in unknown states (along with ancillas) to the state of $k$ logical qubits encoded in a stabiliser code with $n$ physical qubits. Labelling the ancillas in the initial state $k, k+1,\ldots ,n-1$, we note that the initial product state is a $+1$-eigenstate of the stabilisers $Z_k,Z_{k+1},\ldots,Z_{n-1}$. Thus, we wish to find a unitary encoding circuit that maps the stabilisers $Z_k,Z_{k+1},\ldots,Z_{n-1}$ of the product state to a generating set for the stabiliser group $\mathcal{S}$ of the code. The circuit must also map the logical operators $Z_0,Z_1,\ldots,Z_{k-1}$ and  $X_0,X_1,\ldots,X_{k-1}$ of the physical qubits to the corresponding logical operators $\bar{Z}_0,\bar{Z}_1,\ldots,\bar{Z}_{k-1}$ and  $\bar{X}_0,\bar{X}_1,\ldots,\bar{X}_{k-1}$ of the encoded qubits (up to stabilisers).

Applying a unitary $U$ to an eigenstate $\ket{\psi}$ of an operator $S$ (with eigenvalue $s$) gives $US\ket{\psi}=sU\ket{\psi}=USU^\dagger U\ket{\psi}$: an eigenstate of $S$ becomes an eigenstate of $USU^\dagger$. Therefore, we wish to find a unitary encoding circuit that, acting under conjugation, transforms the stabilisers and logicals of the initial product state into the stabilisers and logicals of the encoded state. 

The CNOT gate, acting by conjugation, transforms Pauli $X$ and $Z$ operators as follows:
\begin{align}\label{eq:cnotstabilisers}
    XI \leftrightarrow XX, \quad IZ \leftrightarrow ZZ,
\end{align}
and leaves $ZI$ and $IX$ invariant. Here $\sigma \sigma^\prime$ for $\sigma,\sigma^\prime\in \{I, Z, X\}$ denotes $\sigma_C\otimes \sigma_T$ with $C$ and $T$ the control and target qubit of the CNOT respectively. Since $Z=HXH$ and $X=HZH$, a Hadamard gate $H$ transforms an eigenstate of $Z$ into an eigenstate of $X$ and vice versa. We will show how these relations can be used to generate unitary encoding circuits for the surface code using only CNOT and Hadamard gates.

As an example, consider the problem of generating the encoding circuit for the repetition code, which has stabilisers $Z_0Z_1$ and $Z_1Z_2$. We start in the product state $\ket{\phi}\ket{0}\ket{0}$ which has stabilisers $Z_1$ and $Z_2$. We first apply CNOT$_{01}$ which transforms the stabiliser $Z_1\rightarrow Z_0Z_1$ and leaves $Z_2$ invariant. Then applying CNOT$_{12}$ transforms $Z_2\rightarrow Z_1Z_2$ and leaves $Z_0Z_1$ invariant. We can also verify that the logical $X$ undergoes the required transformation $X_0\rightarrow \bar{X}_0\coloneqq X_0X_1X_2$.
\section{General Encoding Methods for Stabiliser Codes}

There exists a general method for generating an encoding circuit for any stabiliser code~\cite{gottesman1997stabilizer, Cleve_1997}, which we review in Appendix A. The specific structure of the output of this method means it can immediately be rearranged to depth $O(n)$. Using general routing procedures presented in \cite{cheung2007translation, beals2013efficient,brierley2015efficient} the output circuit could be adapted to a surface architecture with overhead $O(\sqrt{n})$, giving a circuit with depth $O(n\sqrt{n})$. This matches the scaling $O(\min(2n^2,4nD\Delta))$ in depth for stabiliser circuits achieved in \cite{wu2019optimization}, where $D$ and $\Delta$ are the diameter and degree respectively of the underlying architecture graph. Any stabiliser circuit has an equivalent skeleton circuit \cite{Maslov_2007}, and so can be implemented on a surface architecture with depth $O(n) = O(L^2)$, matching the previously best known scaling \cite{dennis2002topological} for encoding the planar code. $O(n)$ is an optimal bound on the depth of the set of all stabiliser circuits \cite{Maslov_2007}, so we look beyond general methods and work with the specifics of the planar encoding circuit to improve on \cite{dennis2002topological}. 
\section{Optimal encoder for the planar code}\label{sec:planar_encoding}

Dennis \textit{et al.}~\cite{dennis2002topological} showed how the methods outlined in section \ref{sec:encoding_unknown} can be used to generate an encoding circuit for the planar surface code. The inductive step in their method requires $\Omega(L)$ time steps and encodes a distance $L+1$ planar code from a distance $L$ code by turning smooth edges into rough edges and vice versa. As a result encoding a distance $L$ planar code from an unencoded qubit requires $
\Omega(L^2)$ time steps, which is quadratically slower than the lower bound given by Bravyi \textit{et al.}~\cite{bravyi2006lieb}.

\begin{figure}
    \centering
    \includegraphics[trim={0 0 3cm 0}, width=0.8\columnwidth]{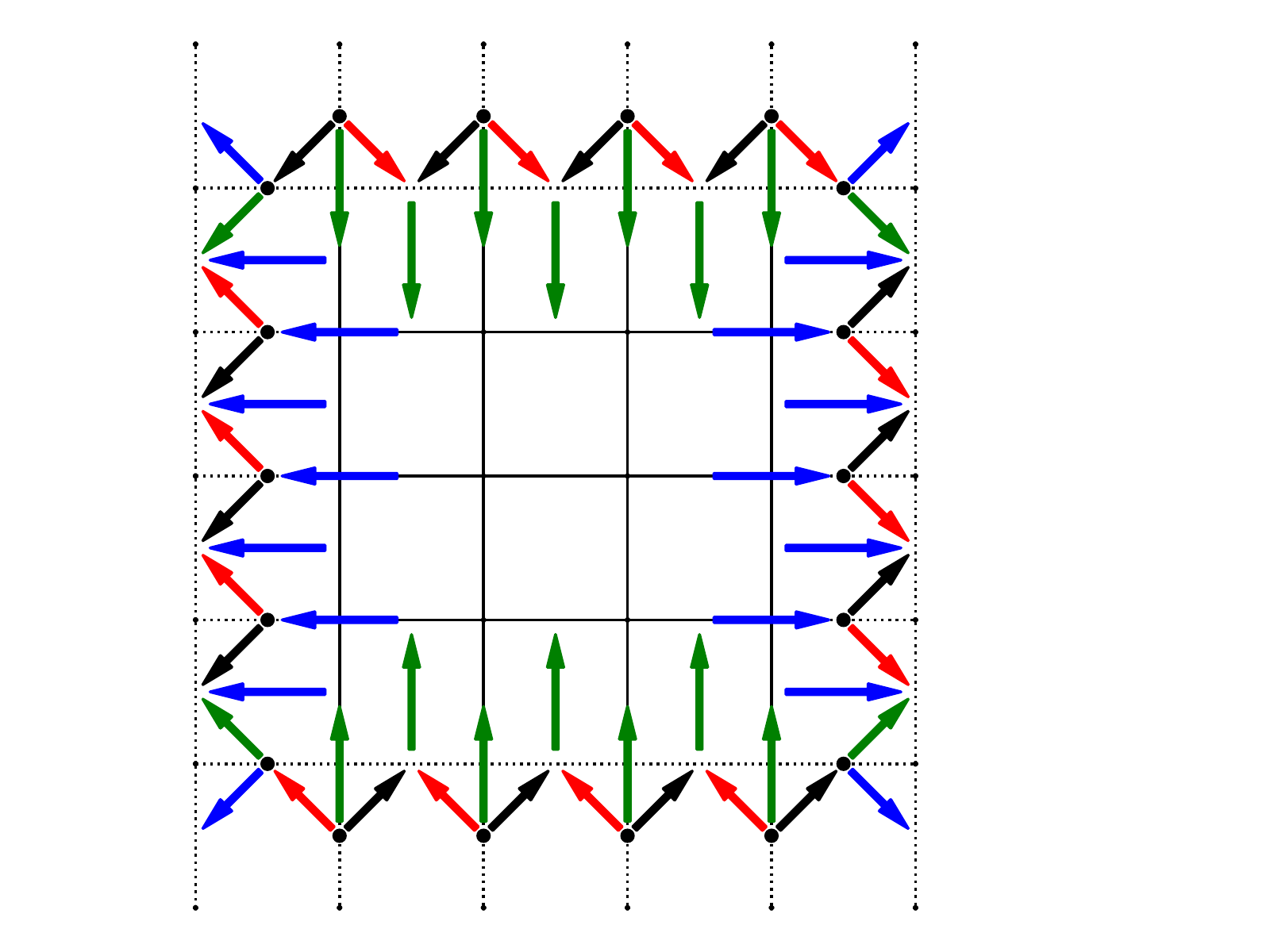}
    \caption{Circuit to encode a distance 6 planar code from a distance 4 planar code. Each edge corresponds to a qubit. Each arrow denotes a CNOT gate, pointing from control to target. Filled black circles (centred on edges) denote Hadamard gates, which are applied at the beginning of the circuit. The colour of each CNOT gate (arrow) denotes the time step in which it is applied. The first, second, third and fourth time steps correspond to the blue, green, red and black CNOT gates respectively. Solid edges correspond to qubits originally encoded in the L=4 planar code, whereas dotted edges correspond to additional qubits that are encoded in the L=6 planar code.}
    \label{fig:planar_L4_to_L6}
\end{figure}

However, here we present a local unitary encoding circuit for the planar code that requires only $2L$ time steps to encode a distance $L$ planar code. The inductive step in our method, shown in \Cref{fig:planar_L4_to_L6} for $L=4$, encodes a distance $L+2$ planar code from a distance $L$ planar code using 4 time steps, and does not rotate the code.  This inductive step can then be used recursively to encode an unencoded qubit into a distance $L$ planar code using $2L$ time steps. If $L$ is odd, the base case used is the distance 3 planar code, which can be encoded in 6 time steps. If $L$ is even, a distance 4 planar code is used as a base case, which can be encoded in 8 time steps. Encoding circuits for the distance 3 and 4 planar codes are given in Appendix~\ref{app:additional_planar_encoders}. Our encoding circuit therefore matches the $\Omega(L)$ lower bound provided by Bravyi \textit{et al.}~\cite{bravyi2006lieb}.

\begin{figure}
\includegraphics{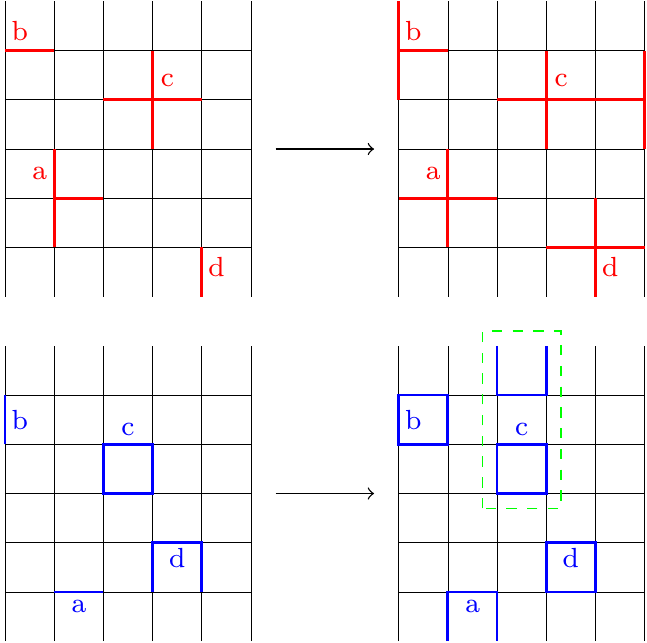}
\caption{The transformation of the stabiliser generators of the $L=4$ planar surface code when the circuit in \Cref{fig:planar_L4_to_L6} is applied. Top: the four main types of site stabilisers acted on nontrivially by the encoding circuit (labelled a-d) are shown in red before (left) and after (right) the encoding circuit is applied. On the left we assume that the ancillas have already been initialised in the $\ket{+}$ state ($H$ applied). Bottom: the four main types of plaquette stabilisers (also labelled a-d) are shown in blue before (left) and after (right) the encoding circuit is applied. Plaquette c has two connected components after the circuit is applied (right), and is enclosed by a green dashed line for clarity.}
\label{fig:planar_generators_mapped}
\end{figure}

Since the circuit for the inductive step in \Cref{fig:planar_L4_to_L6} uses only CNOT and $H$ gates, we can verify its correctness by checking that stabiliser generators and logicals of the distance $L$ surface code are mapped to stabiliser generators and logicals of the distance $L+2$ surface code using the conjugation rules explained in \Cref{sec:encoding_unknown}.
We show how each type of site and plaquette stabiliser generator is mapped by the inductive step of the encoding circuit in \Cref{fig:planar_generators_mapped}.
Note that the site stabiliser generator labelled c (red) is mapped to a weight 7 stabiliser in the $L=6$ planar code: this is still a valid generator of stabiliser group, and the standard weight four generator can be obtained by multiplication with a site of type b.
Similarly, the plaquette stabiliser generator labelled c becomes weight 7, but a weight four generator is recovered from multiplication by a plaquette of type a.
Therefore, the stabiliser group of the $L=4$ planar code is mapped correctly to that of the $L=6$ planar code, even though minimum-weight generators are not mapped explicitly to minimum-weight generators.
Using \Cref{eq:cnotstabilisers} it is straightforward to verify that the $X$ and $Z$ logical operators of the $L=4$ planar code are also mapped to the $X$ and $Z$ logicals of the $L=6$ planar code by the inductive step.

We can also encode rectangular planar codes with height $H$ and width $W$ by first encoding a distance $\min(H,W)$ square planar code and then using a subset of the gates in \Cref{fig:planar_L4_to_L6} (given explicitly in Appendix~\ref{app:additional_planar_encoders}) to either increase the width or the height as required. Increasing either the width or height by two requires three time steps, therefore encoding a $H\times W$ rectangular planar code from an unencoded qubit requires $2\min(H,W)+3\left\lceil\frac{|H-W|}{2}\right\rceil$ time steps.

In Appendix~\ref{app:rotated_code} we also provide an optimal encoder for the \textit{rotated} surface code, which uses fewer physical qubits for a given distance $L$~\cite{bombin2007optimal}. Our encoding circuit also uses an inductive step that increases the distance by two using four time steps, and therefore uses $2L + O(1)$ time steps to encode a distance $L$ rotated surface code.

\section{Local Renormalisation Encoder for the Toric Code}\label{sec:RenormalisationEncoder}
In this section we will describe an $O(L)$ encoder for the toric code based on the multi-scale entanglement renormalisation ansatz (MERA). The core of this method is to enforce locality in the Renormalisation Group (RG) encoder given by Aguado and Vidal~\cite{aguado_MERA}. The RG encoder starts from an $L=2$ toric code and then uses an $O(1)$ depth inductive step which enlarges a distance $2^k$ code to a distance $2^{k+1}$ code, as shown in \Cref{fig:L2_to_L4} for the first step ($k=1$) (and reviewed in more detail in Appendix~\ref{app:renormalisation_group}). The $L=2$ base case toric code can be encoded using the method given by Gottesman in Ref.~\cite{gottesman1997stabilizer}, as shown in Appendix \ref{basetoric}. While the RG encoder takes $O(\log L)$ time, it is non-local in it's original form.

\begin{figure}
    \centering
    \subfloat[\label{fig:timing1}]{
        \centering
        \includegraphics[width=0.21\textwidth]{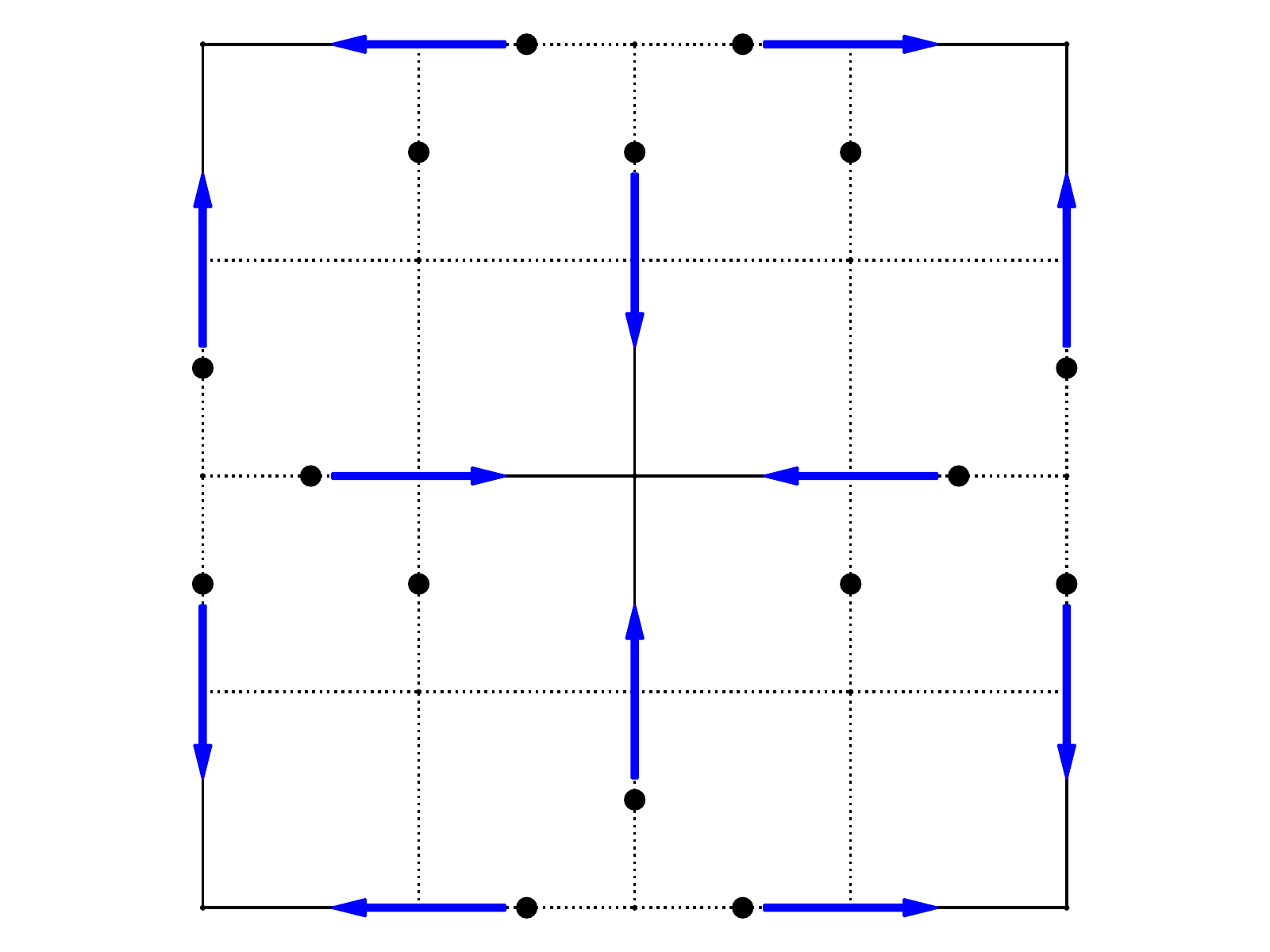}  
     }
    \hfill
    \subfloat[\label{fig:timing2}]{
        \centering
        \includegraphics[width=0.21\textwidth]{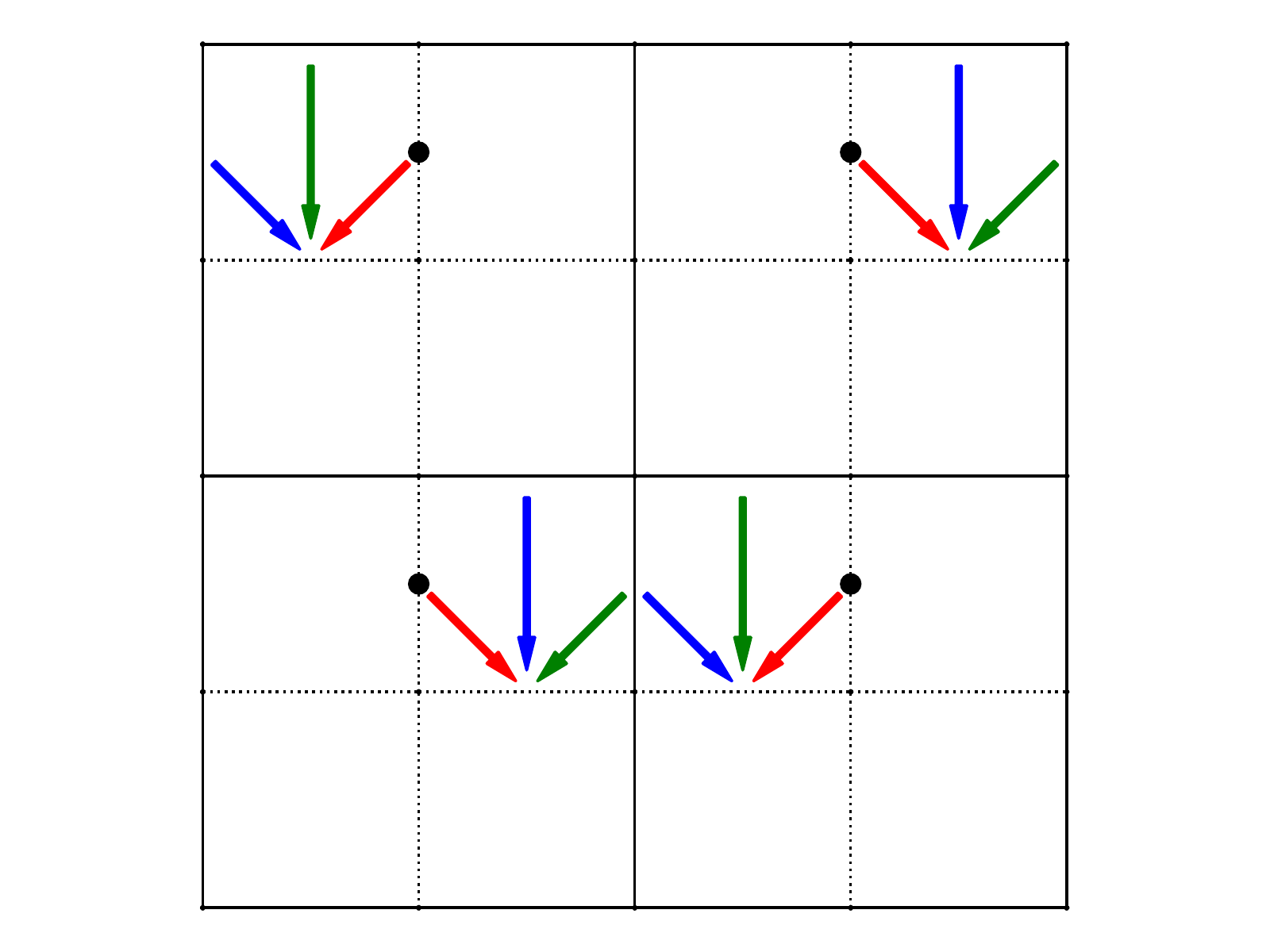} 
     }
    \hfill
    \subfloat[\label{fig:timing3}]{
        \centering
        \includegraphics[width=0.21\textwidth]{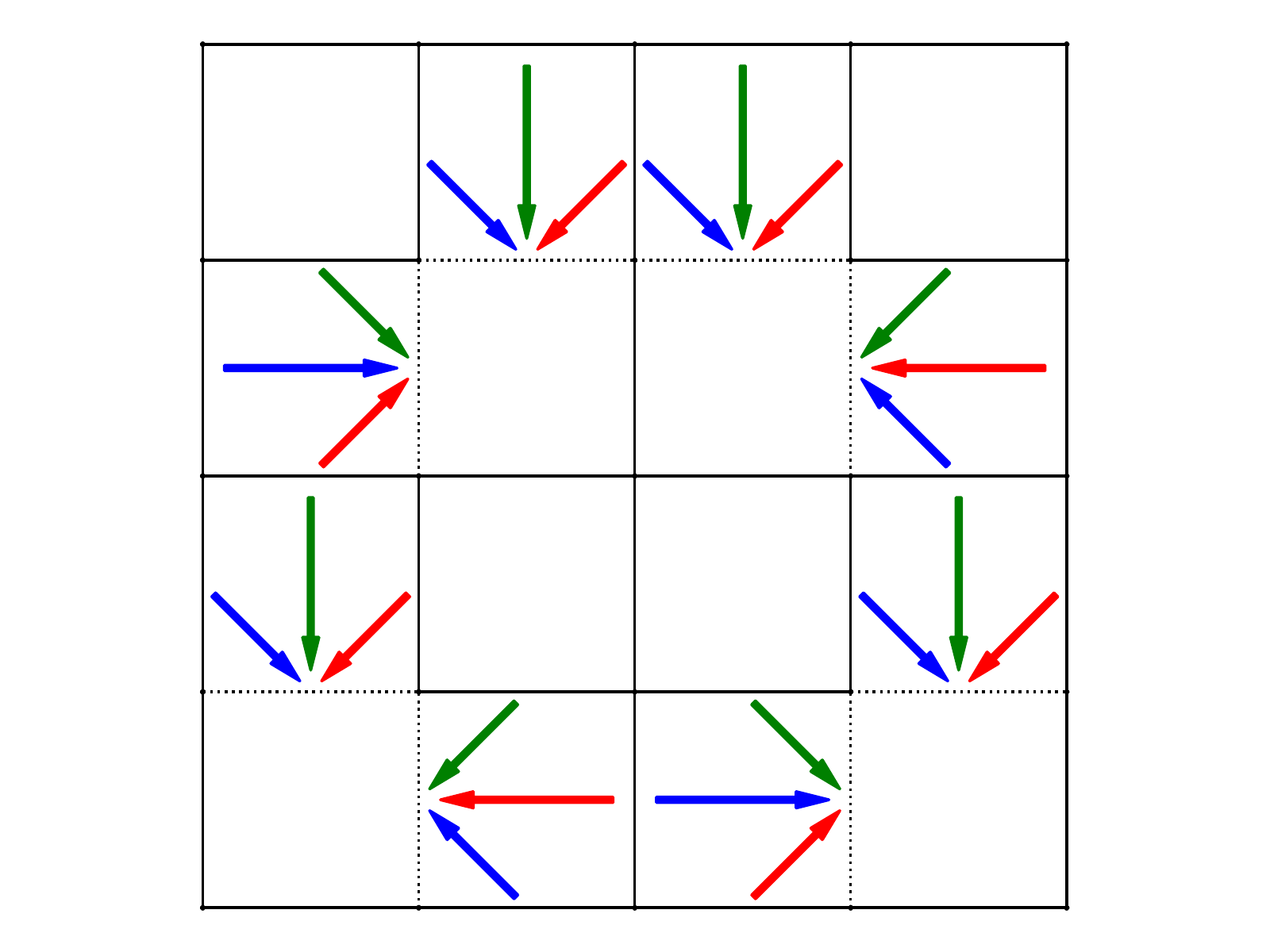} 
    }
    \caption{Encoding a distance 4 toric code from a distance 2 toric code using the Renormalisation Group encoder of Aguado and Vidal~\cite{aguado_MERA}. Dashed edges, dashed edges with a node and solid edges correspond to decoupled ancillae in $\ket{0}$, in $\ket{+}$, and to qubits entangled with the original code respectively. Opposite edges are identified. Arrows denote CNOT operations from control to target qubits, and monochromatic gates in stages (b) and (c) may be executed in a single timestep.}
    \label{fig:L2_to_L4}
\end{figure}

In order to enforce locality in the RG encoder, we wish to find an equivalent circuit that implements an identical operation on the same input state, using quantum gates that act locally on the physical architecture corresponding to the final distance $L$ toric code (here a gate is \textit{local} if it acts only on qubits that belong to either the same site or plaquette). One approach to enforce locality in a quantum circuit is to insert SWAP gates into the circuit to move qubits adjacent to each other where necessary. Any time step of a quantum circuit can be made local on a $L\times L$ 2D nearest-neighbour (2DNN) grid architecture using at most $O(L)$ time steps, leading to at most a multiplicative $O(L)$ overhead from enforcing locality~\cite{cheung2007translation, beals2013efficient, brierley2015efficient}. Placing an ancilla in the centre of each site and plaquette, we see that the connectivity graph of our physical architecture has a 2DNN grid as a subgraph. Therefore, using SWAP gates to enforce locality in the RG encoder immediately gives us a $O(L\log L)$ local unitary encoding circuit for the toric code which, while an improvement on the $O(L^2)$ encoder in Ref.~\cite{dennis2002topological}, does not match the $\Omega(L)$ lower bound.

However, we can achieve $O(L)$ complexity by first noticing that all `quantum circuit' qubits which are acted on non-trivially in the first $k$ steps of the RG encoder can be mapped to physical qubits in a $2^{k+1}\times 2^{k+1}$ square region of the physical architecture. Therefore, the required operations in iteration $k$ can all be applied within a $2^{k+1}\times 2^{k+1}$ region that also encloses the regions used in the previous steps. In Appendix \ref{routing} we use this property to provide circuits for routing quantum information using SWAP gates (and no ancillas) that enforce locality in each of the $O(1)$ time steps in iteration $k$ using $O(2^{k+1})$ time steps. This leads to a total complexity of $\sum_{k=1}^{\log_2(L)-1} O(2^{k+1}) = O(L)$ for encoding a distance $L$ code, also achieving the lower bound given by Bravyi \textit{et al.}~\cite{bravyi2006lieb}. In Appendix~\ref{routing} we provide a more detailed analysis to show that the total time complexity is $15L/2 - 6\log_2 L + 7 \sim O(L)$. Unlike the other encoders in this paper (which work for all $L$), the RG encoder clearly can only be applied when $L$ is a power of 2.

\begin{figure}
    \centering
    \includegraphics[trim={0 0 0 0}, width=0.8\columnwidth]{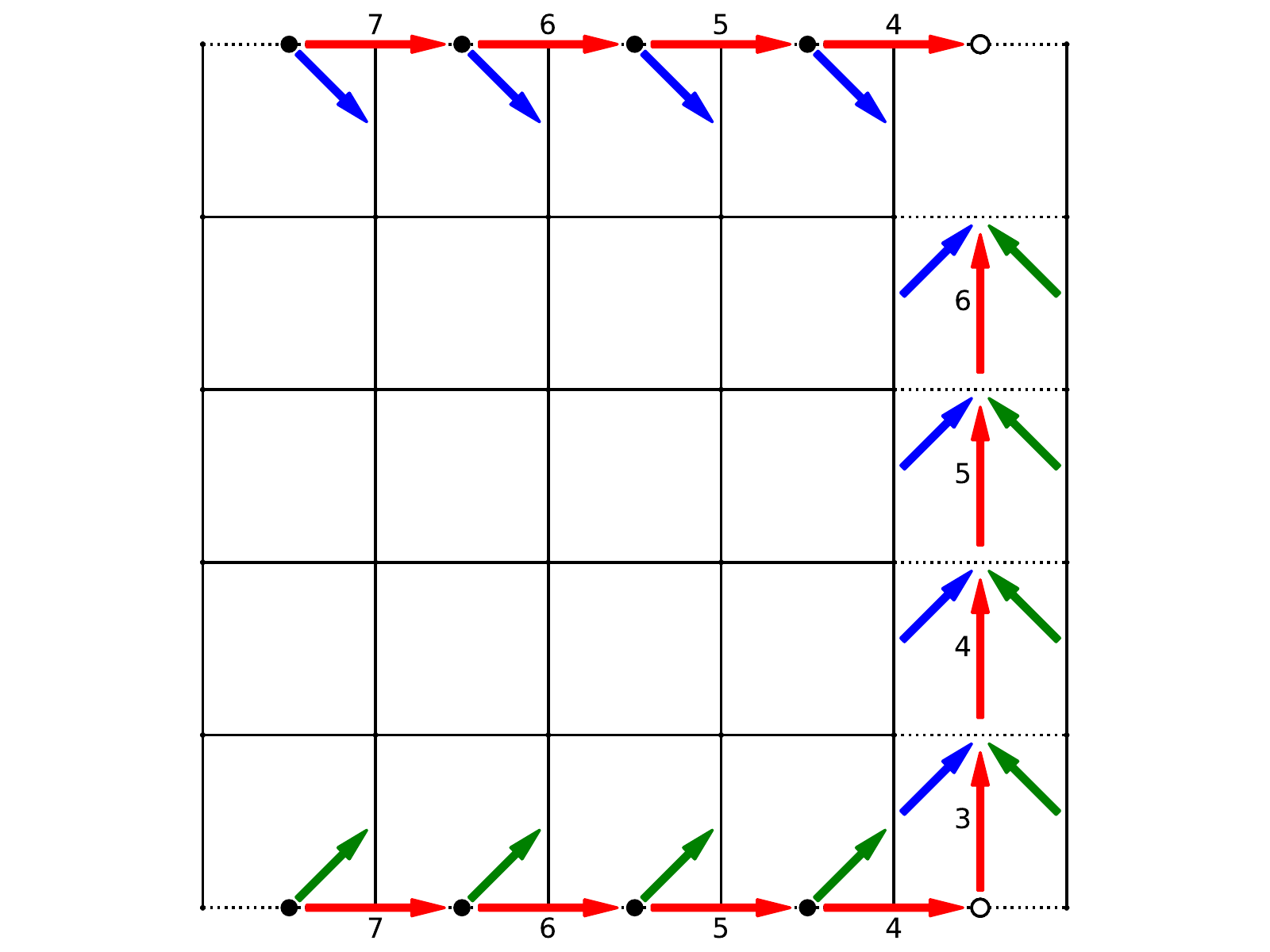}
    \caption{Circuit to encode a distance 5 toric code from a distance 5 planar code. Solid edges correspond to qubits in the original planar code and dotted edges correspond to qubits added for the toric code. Opposite edges are identified. Arrows denote CNOT gates, and filled black circles denote Hadamard gates applied at the beginning of the circuit. Blue and green CNOT gates correspond to those applied in the first and second time step respectively. Red CNOTs are applied in the time step that they are numbered with. The hollow circles denote the unencoded qubit that is to be encoded into the toric code.}
    \label{fig:planar_to_toric}
\end{figure}

\section{Encoding a toric code from a planar code}

While the method in section \ref{sec:planar_encoding} is only suitable for encoding planar codes, we will now show how we can encode a distance $L$ toric code from a distance $L$ planar code using only local unitary operations. Starting with a distance $L$ planar code, $2(L-1)$ ancillas each in a $\ket{0}$ state, and an additional unencoded logical qubit, the circuit in \Cref{fig:planar_to_toric} encodes a distance $L$ toric code using $L+2$ time steps. 
The correctness of this step can be verified using \Cref{eq:cnotstabilisers}: each ancilla initialised as $\ket{0}$ (stabilised by $Z$) is mapped to a plaquette present in the toric code but not the planar code. 
Likewise, each ancilla initialised in $\ket{+}$ using an $H$ gate (stabilised by $X$) is mapped to a site generator in the toric code but not the planar code.
The weight-three site and plaquette stabilisers on the boundary of the planar code are also mapped to weight four stabilisers in the toric code.
Finally, we see that $X$ and $Z$ operators for the unencoded qubit (the hollow circle in \Cref{fig:planar_to_toric}) are mapped to the second pair of $X$ and $Z$ logicals in the toric code by the circuit, leaving the other pair of $X$ and $Z$ logicals already present from the planar code unaffected.

Therefore, encoding two unencoded qubits in a toric code can be achieved using $3L+2$ time steps using the circuits given in this section and in section \ref{sec:planar_encoding}. Similarly, we can encode a planar code using the local RG encoder for the toric code, before applying the inverse of the circuit in \Cref{fig:planar_to_toric}.

\section{Encoding a 3D Surface Code}

\begin{figure}
    \centering
    \subfloat[]{
    \includegraphics[width=0.3\columnwidth,trim={0 0 9cm 0}]{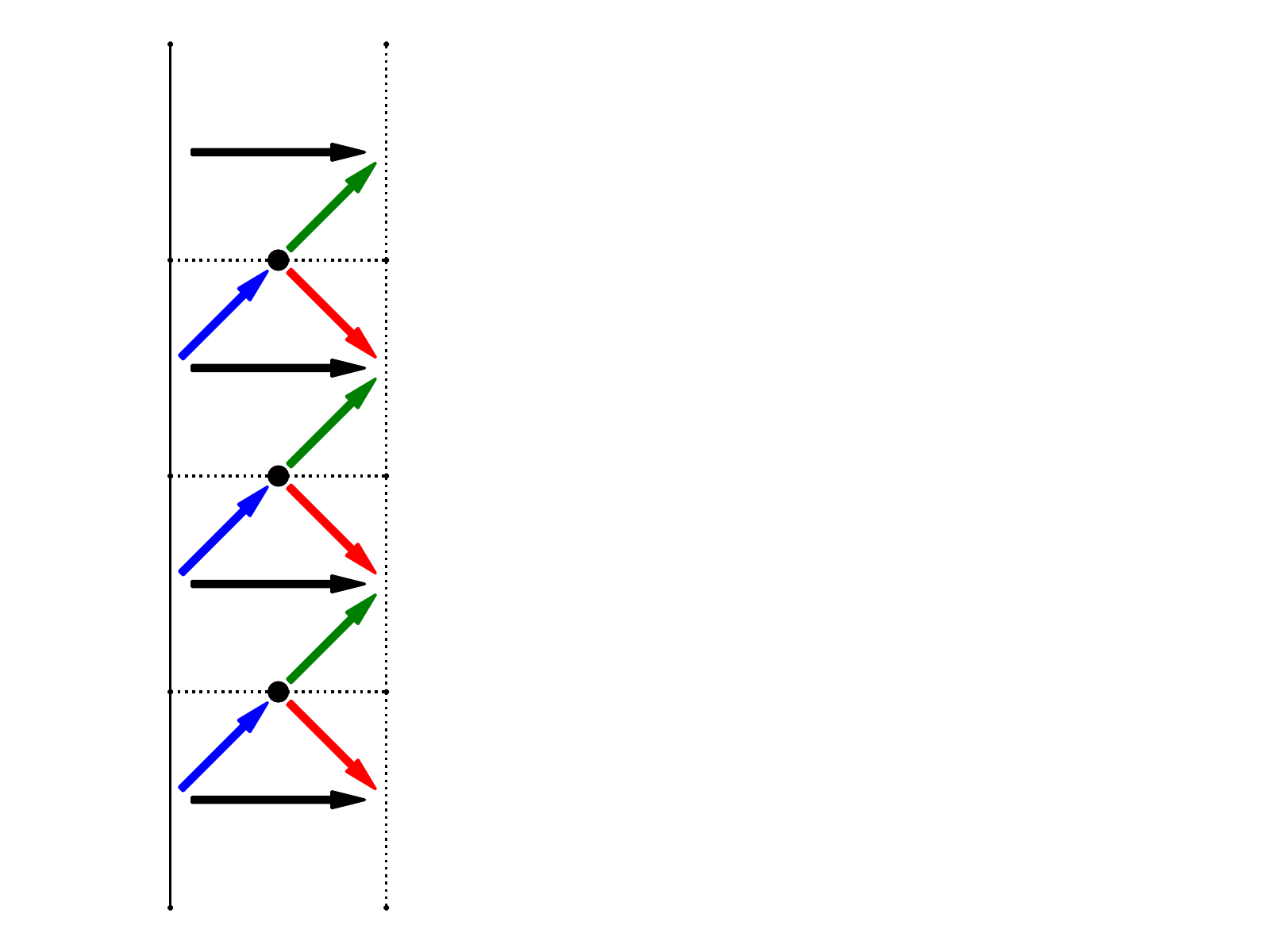}
    }
    \hfill
    \subfloat[]{
    \includegraphics[width=0.45\columnwidth, trim={2cm 0 2cm 0}]{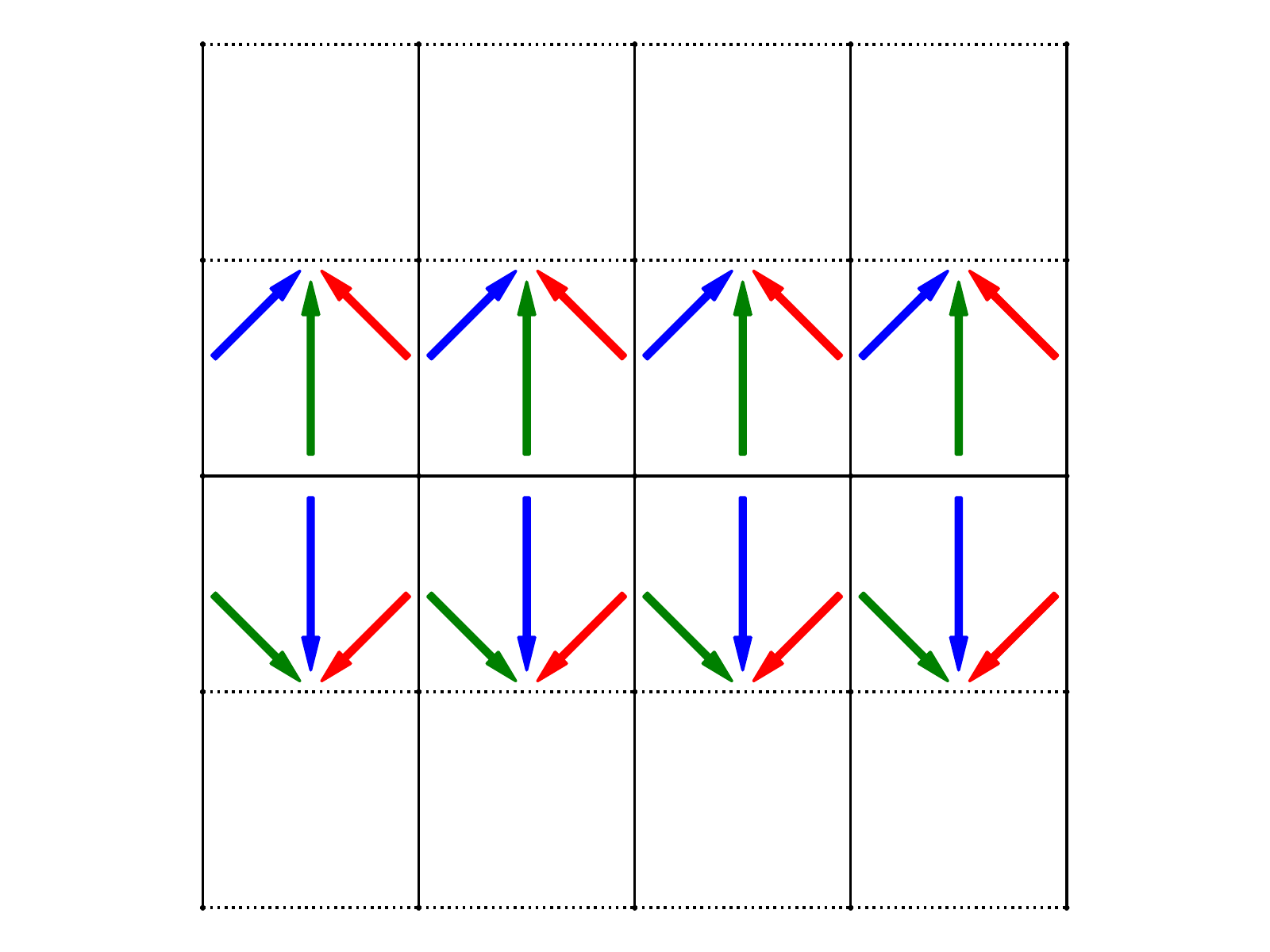}
    }
    \caption{(a) Circuit to encode a $4\times 2$ planar code from a four qubit repetition code (where adjacent qubits in the repetition code are stabilised by $XX$). Applied to a column of qubits corresponding to a surface code $\bar{Z}$, this encodes a layer in the $yz$-plane of a 3D surface code. (b) Circuit to encode the $xz$-plane of a 3D surface code once the $yz$-plane layers and a layer in the $xy$-plane have been encoded. Arrows denote CNOT gates pointing from control to target, and blue, green, red and black CNOT gates correspond to the first, second, third and fourth time steps respectively. Solid and dotted edges correspond to qubits that are initially entangled and in a product state respectively.}
    \label{fig:3D_surface_code}
\end{figure}

We will now show how the techniques developed to encode a 2D planar code can be used to encode a distance $L$ 3D surface code using $O(L)$ time steps. We first encode a distance $L$ planar code using the method given in section \ref{sec:planar_encoding}. This planar code now forms a single layer in the $xy$-plane of a 3D surface code (where the $y$-axis is defined to be aligned with a $Z$-logical in the original planar code). Using the circuit given in \Cref{fig:3D_surface_code}(a), we encode each column of qubits corresponding to a $Z$ logical in the planar code into a layer of the 3D surface code in the $yz$-plane (which has the same stabiliser structure as a planar code if the rest of the $x$-axis is excluded). Since each layer in the $yz$-plane can be encoded in parallel, this stage can also be done in $O(L)$ time steps. If we encode each layer in the $yz$-plane such that the original planar code intersects the middle of each layer in the $yz$-plane, then each layer in the $xz$-plane now has the stabiliser structure shown in \Cref{fig:3D_surface_code}(b). Using the circuit in \Cref{fig:3D_surface_code}(b) repeatedly, all layers in the $xz$-plane can be encoded in parallel in $O(L)$ time steps. Therefore, a single unknown qubit can be encoded into a distance $L$ 3D surface code in $O(L)$ time steps.

\section{Encoding circuit for the compact mapping}

Fermion to qubit mappings are essential for simulating fermionic systems using quantum computers, and an encoding circuit for such a mapping is an important subroutine in many quantum simulation algorithms. 
We now show how we can use our encoding circuits for the surface code to construct encoding circuits that prepare fermionic states in the compact mapping~\cite{derby2021compact}, a fermion to qubit mapping that is especially efficient for simulating the Fermi-Hubbard model.
A fermion to qubit mapping defines a representation of fermionic states in qubits, as well as a representation of each fermionic operator in terms of Pauli operators.
Using such a mapping, we can represent a fermionic Hamiltonian as a linear combination $H=\sum_i \alpha_i P_i$ of tensor products of Pauli operators $P_i$, where $\alpha_i$ are real coefficients.
We can then simulate time evolution $e^{-iHt}$ of $H$ (e.g.~using a Trotter decomposition), which can be used in the quantum phase estimation algorithm to determine the eigenvalues of $H$.
The mapped Hamiltonian $H$ can also be used in the variational quantum eigensolver algorithm (VQE), where we can estimate the energy $\bra{\psi}H\ket{\psi}$ of a trial state $\ket{\psi}$ by measuring each Pauli term $\bra{\psi}P_i\ket{\psi}$ individually.

The Jordan-Wigner (JW) transformation maps fermionic creation ($a_i^\dagger$) and annihilation ($a_i$) operators to qubit operators in such a way that the canonical fermionic anti-commutation relations
\begin{equation}
\{a_{i}^{\dagger}, a_{j}^{\dagger}\}=0,\left\{a_{i}, a_{j}\right\}=0,\{a_{i}^{\dagger}, a_{j}\}=\delta_{i j}
\end{equation}
are satisfied by the encoded qubit operators.
The qubit operators used to represent $a_i^\dagger$ and $a_i$ are
\begin{align}
a_i^\dagger &\rightarrow Z_1\ldots Z_{i-1}\sigma_i^+ \\
a_i &\rightarrow Z_1\ldots Z_{i-1}\sigma_i^-
\end{align}
where $\sigma^+\coloneqq (X_i-iY_i)/2$ and $\sigma^-\coloneqq (X_i+iY_i)/2$.
Each electronic basis state (with $m$ modes) in the JW transformation is represented by $m$ qubits simply as a computational basis state $\ket{\omega_1,\omega_2,\ldots,\omega_m}$ where $\omega_i=1$ or $\omega_i=0$ indicates that mode $i$ is occupied or unoccupied by a fermion, respectfully.

A drawback of the Jordan-Wigner transformation is that, even if a fermionic operator acts on $O(1)$ modes, the corresponding JW-mapped qubit operator can still act on up to $O(m)$ qubits.
When mapped qubit operators have larger weight, the depth and number of gates required to simulate time evolution of a mapped Hamiltonian also tend to increase, motivating the design of fermion-to-qubit mappings that map fermionic operators to qubit operators that are both low weight and geometrically local.

Several methods have been proposed for mapping geometrically local fermionic operators to geometrically local qubit operators~\cite{verstraete2005mapping,whitfield2016local,steudtner2019quantum,bravyi2002fermionic,setia2019superfast,jiang2019majorana,derby2021compact}, all of which introduce auxiliary qubits and encode fermionic Fock space into a subspace of the full $n$-qubit system, defined as the $+1$-eigenspace of elements of a stabiliser group $\mathcal{S}$.
Mappings that have this property as referred to as \textit{local}.

We will now focus our attention on a specific local mapping, the compact mapping~\cite{derby2021compact}, since its stabiliser group is very similar to that of the surface code.
As we will see, this close connection to the surface code allows us to use the encoding circuits we have constructed for the surface code to encode fermionic states in the compact mapping.
The compact mapping maps nearest-neighbour hopping ($a_i^\dagger a_j + a_j^\dagger a_i$) and Coulomb ($a_i^\dagger a_i a_j^\dagger a_j$) terms to Pauli operators with weight at most 3 and 2, respectfully, and requires 1.5 qubits for each fermionic mode~\cite{derby2021compact}.
Rather than mapping individual fermionic creation and annihilation operators, the compact mapping instead defines a representation of the fermionic edge ($E_{j k}$) and vertex ($V_{j}$) operators, defined as
\begin{equation}
E_{j k}\coloneqq-i \gamma_{j} \gamma_{k}, \quad V_{j}\coloneqq-i \gamma_{j} \bar{\gamma}_{j},
\end{equation}
where $\gamma_j\coloneqq a_j+a_j^\dagger$ and $\bar{\gamma}_j\coloneqq (a_j-a_j^\dagger)/i$ are Majorana operators.
The vertex and edge operators must satisfy the relations
\begin{equation}
\left[E_{i j}, V_{l}\right]=0, \quad\left[V_{i}, V_{j}\right]=0, \quad\left[E_{i j}, E_{l n}\right]=0.
\end{equation}
for all $i\neq j \neq l \neq n$, and
\begin{equation}
\left\{E_{i j}, E_{j k}\right\}=0, \quad\left\{E_{j k}, V_{j}\right\}=0.
\end{equation}
In the compact mapping, there is a ``primary'' qubit associated with each of the $m$ fermionic modes, and there are also $m/2$ ``auxiliary'' qubits. 
Each vertex operator $V_j$ is mapped to the Pauli operator $Z_j$ on the corresponding primary qubit.
We denote the mapped vertex and edge operators by $\tilde{V}_j$ and $\tilde{E}_{ij}$, respectfully, and so we have $\tilde{V}_j\coloneqq Z_j$.
Each edge operator $E_{ij}$ is mapped (up to a phase factor) to a three-qubit Pauli operator of the form $XYX$ or $XYY$, with support on two vertex qubits and a neighbouring ``face'' qubit.
The precise definition of the edge operators is not important for our purposes, and we refer the reader to Ref.~\cite{derby2021compact} for details.

The vertex and edge operators define a graph (in which they correspond to vertices and edges, respectfully), and an additional relation that must be satisfied in the mapping is that the product of any loop of edge operators must equal the identity:
\begin{equation}\label{eq:identity_loop}
i^{(|p|-1)}\prod_{i=1}^{(|p|-1)}\tilde{E}_{p_i p_{i+1}}=1,
\end{equation}
where here $p=\{p_1,p_2,\ldots\}$ is a sequence of vertices along any cycle in the graph.
The relation of \Cref{eq:identity_loop} can be satisfied by ensuring that the qubit operator corresponding to any mapped loop of edge operators is a stabiliser, if it is not already trivial, thereby ensuring that the relations are satisfied within the $+1$-eigenspace of the stabilisers.

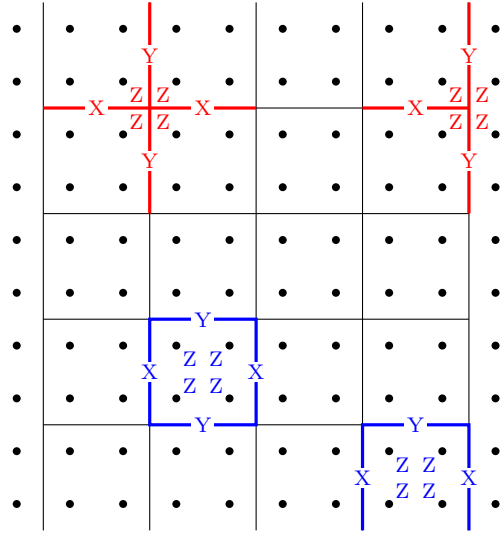
\begin{figure}
\begin{tikzpicture}[scale=0.35]
\foreach \i in {0,1,2,3,4}
{
\draw (4*\i+1,1) -- (4*\i+1, 21);
}
\foreach \i in {1,2,3,4}
{
\draw (1,4*\i+1) -- (17, 4*\i+1);
}
\foreach \x in {0,...,9}
	\foreach \y in {1,...,10}
		{
			\node[circle,fill=black,inner sep=0pt,minimum size=3pt] (a) at (2*\x,2*\y) {};
		}
\draw[very thick, blue] (5,5) rectangle (9,9);
\node[blue, fill=white, inner sep=1] at (7,9) {\footnotesize Y};
\node[blue, fill=white, inner sep=1] at (9,7) {\footnotesize X};
\node[blue, fill=white, inner sep=1] at (5,7) {\footnotesize X};
\node[blue, fill=white, inner sep=1] at (7,5) {\footnotesize Y};
\node[blue] at (6.5,6.5) {\footnotesize Z};
\node[blue] at (7.5,6.5) {\footnotesize Z};
\node[blue] at (6.5,7.5) {\footnotesize Z};
\node[blue] at (7.5,7.5) {\footnotesize Z};
\draw[very thick, blue] (13,1) -- (13,5) -- (17,5) -- (17,1);
\node[blue, fill=white, inner sep=1] at (15,5) {\footnotesize Y};
\node[blue, fill=white, inner sep=1] at (13,3) {\footnotesize X};
\node[blue, fill=white, inner sep=1] at (17,3) {\footnotesize X};
\node[blue] at (14.5,2.5) {\footnotesize Z};
\node[blue] at (15.5,2.5) {\footnotesize Z};
\node[blue] at (14.5,3.5) {\footnotesize Z};
\node[blue] at (15.5,3.5) {\footnotesize Z};
\draw[very thick, red] (5,13) -- (5,21);
\draw[very thick, red] (1,17) -- (9,17);
\node[red, fill=white, inner sep=1] at (5,19) {\footnotesize Y};
\node[red, fill=white, inner sep=1] at (3,17) {\footnotesize X};
\node[red, fill=white, inner sep=1] at (7,17) {\footnotesize X};
\node[red, fill=white, inner sep=1] at (5,15) {\footnotesize Y};
\node[red] at (4.5,16.5) {\footnotesize Z};
\node[red] at (5.5,16.5) {\footnotesize Z};
\node[red] at (4.5,17.5) {\footnotesize Z};
\node[red] at (5.5,17.5) {\footnotesize Z};
\draw[very thick, red] (13,17) -- (17,17);
\draw[very thick, red] (17,13) -- (17,21);
\node[red, fill=white, inner sep=1] at (15,17) {\footnotesize X};
\node[red, fill=white, inner sep=1] at (17,15) {\footnotesize Y};
\node[red, fill=white, inner sep=1] at (17,19) {\footnotesize Y};
\node[red] at (16.5,16.5) {\footnotesize Z};
\node[red] at (17.5,16.5) {\footnotesize Z};
\node[red] at (16.5,17.5) {\footnotesize Z};
\node[red] at (17.5,17.5) {\footnotesize Z};
\end{tikzpicture}
\caption{The stabilisers of the compact mapping. A primary qubit is associated with each black circle, and an auxiliary qubit is associated with each edge of the surface code lattice. There is a plaquette stabiliser (blue) associated with each face of the surface code lattice, acting as $YXXY$ on the edges adjacent to the face, and as $Z$ on each of the four closest primary qubits. There is also a site stabiliser (red) associated with each vertex of the surface code lattice, also acting as $YXXY$ on the edges adjacent to the vertex, and as $Z$ on each of the four closest primary qubits.}
\label{fig:compact_mapping_stabilisers}
\end{figure}

The stabiliser group $\mathcal{S}$ of the compact mapping is therefore defined by \Cref{eq:identity_loop} and the definition of each $\tilde{E}_{ij}$.
The $+1$-eigenspace of $\mathcal{S}$ has dimension $2^{m+\Delta}$, where $m$ is the number of modes and $\Delta\in\{-1,0,1\}$ is the \textit{disparity}, which depends on the boundary conditions chosen for the square lattice geometry.
We will only consider the case where $\Delta=1$, since this choice results in a stabiliser structure most similar to the surface code.
In this $\Delta=1$ case the full Fock space is encoded, along with a topologically protected logical qubit.
The stabilisers of the compact mapping (for the case $\Delta=1$) are shown in \Cref{fig:compact_mapping_stabilisers}, from which it is clear that the stabiliser group is very similar to that of the planar surface code, a connection which was first discussed in Ref.~\cite{derby2021compact}.
Indeed, if we consider the support of the stabilisers on only the auxiliary qubits (associated with the edges of the surface code lattice shown in \Cref{fig:compact_mapping_stabilisers}), we recover the stabiliser group of the planar surface code up to single-qubit Clifford gates acting on each qubit.

Using this insight, we can use our surface code encoding circuit to construct a local unitary encoding circuit that prepares a Slater determinant state in the compact mapping, which is often required for its use in quantum simulation algorithms.
Note that we can write each fermionic occupation operator $a_j^\dagger a_j$ for mode $j$ in terms of the corresponding vertex operator $V_j$ as $a_j^\dagger a_j=(I-V_j)/2$, where $I$ is the identity operator.
A Slater determinant state $\ket{\phi_{\mathrm{det}}}$ is then a joint eigenstate of the stabilisers and vertex operators:
\begin{align}
S_i\ket{\phi_{\mathrm{det}}}&=\ket{\phi_{\mathrm{det}}},\quad\forall S_i\in\mathcal{S}, \label{eq:determinant_stabilisers} \\
\tilde{V}_j\ket{\phi_{\mathrm{det}}}&=v_j\ket{\phi_{\mathrm{det}}},\quad\forall \tilde{V}_j\in \tilde{V}, \label{eq:vertex_op_eigenvalues}
\end{align}
where $\mathcal{S}$ is the stabiliser group of the mapping, $\tilde{V}$ is the set of mapped vertex operators, and $v_j\in\{+1,-1\}$ indicates whether mode $j$ is occupied (-1) or unoccupied (+1)~\cite{jiang2019majorana}. 

Let us denote the set of generators of $\mathcal{S}$ defined by the sites and plaquettes in \Cref{fig:compact_mapping_stabilisers} by $\{s_1,s_2,\ldots,s_r\}$ (i.e.~$\mathcal{S}=\langle s_1,s_2,\ldots,s_r \rangle$).
For any Pauli operator $c$, we denote its component acting only on the primary qubits as $c^p$, and its component acting only on auxiliary qubits is denoted $c^a$.
With this notation we can decompose each stabiliser generator as $s_i=s_i^p\otimes s_i^a$, where $|s_i^p|=|s_i^a|=4$ in the bulk of the lattice.
For the compact mapping, where $\tilde{V}_j\coloneqq Z_j$, from \Cref{eq:vertex_op_eigenvalues} we see that the primary qubits are in a product state for all Slater determinant states, and so we can write the state of the system on all qubits as $\ket{\phi}=\ket{\phi}_{p}\otimes\ket{\phi}_{a}$, where $\ket{\phi}_{p}$ is the state of the primary qubits and $\ket{\phi}_{a}$ is the state of the auxiliary qubits.

Our circuit to prepare a Slater determinant state in the compact mapping then proceeds in three steps.
In step one we prepare each primary qubit in state $\ket{0}$ or $\ket{1}$ if the corresponding fermionic mode is unoccupied or occupied, respectfully. 
This ensures that the state satisfies \Cref{eq:vertex_op_eigenvalues} as required, and we denote the resultant state on the primary qubits by $\ket{\phi_{det}}_{p}$.
It now remains to show how we can prepare the state on the auxiliary qubits such that \Cref{eq:determinant_stabilisers} is also satisfied.

In step 2, we prepare a state $\ket{\phi_{surf}}_{a}$ on the auxiliary qubits that is in the $+1$-eigenspace of each stabiliser generator restricted to its support only on the auxiliary qubits.
In other words we prepare the state $\ket{\phi_{surf}}_{a}$ satisfying
\begin{equation}
S_i\ket{\phi_{surf}}_{a}=\ket{\phi_{surf}}_{a}\quad\forall S_i\in\mathcal{S}',
\end{equation}
where $\mathcal{S}'\coloneqq \langle s_1^a, s_2^a,\ldots,s_r^a\rangle$.
The generators of $\mathcal{S}'$ are the same as those of the planar surface code up to local Clifford gates, and so we can prepare $\ket{\phi_{surf}}_{a}$ by encoding the planar surface code on the auxiliary qubits using the circuit from \Cref{sec:planar_encoding} and applying $U_V$ ($U_H$) to each vertical (horizontal) edge of the lattice in \Cref{fig:compact_mapping_stabilisers}, where
\begin{align}
U_V &\coloneqq XHS=\frac{1}{\sqrt{2}}\left(\begin{array}{rr}
1 & -i \\
1 & i
\end{array}\right), \\
U_H &\coloneqq XHSH=\frac{1}{2}\left(\begin{array}{rr}
1-i & 1+i \\
1+i & 1-i
\end{array}\right).
\end{align}
This step can be verified by noticing that, under conjugation, $U_V$ maps $X\rightarrow Z$ and $Y\rightarrow X$, and $U_H$ maps $Y\rightarrow Z$ and $X\rightarrow X$, and so the generators of the surface code (\Cref{fig:surface_code}) are mapped to generators of $\mathcal{S}'$.

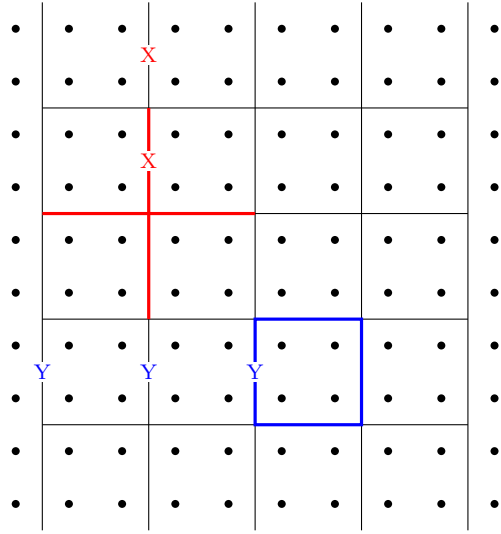
\begin{figure}
\begin{tikzpicture}[scale=0.35]
\foreach \i in {0,1,2,3,4}
{
\draw (4*\i+1,1) -- (4*\i+1, 21);
}
\foreach \i in {1,2,3,4}
{
\draw (1,4*\i+1) -- (17, 4*\i+1);
}
\foreach \x in {0,...,9}
	\foreach \y in {1,...,10}
		{
			\node[circle,fill=black,inner sep=0pt,minimum size=3pt] (a) at (2*\x,2*\y) {};
		}
\draw[very thick, blue] (9,5) rectangle (13,9);
\node[blue, fill=white, inner sep=1] at (9,7) {\footnotesize Y};
\node[blue, fill=white, inner sep=1] at (5,7) {\footnotesize Y};
\node[blue, fill=white, inner sep=1] at (1,7) {\footnotesize Y};
\draw[very thick, red] (1,13) -- (9,13);
\draw[very thick, red] (5,9) -- (5,17);
\node[red, fill=white, inner sep=1] at (5,15) {\footnotesize X};
\node[red, fill=white, inner sep=1] at (5,19) {\footnotesize X};
\end{tikzpicture}
\caption{A $-1$ syndrome on any individual plaquette stabiliser (blue) can be generated by a string of Pauli $Y$ operators (labelled in blue) on qubits on vertical edges joining it to the left (or right) boundary. Similarly, a $-1$ syndrome on any site stabiliser (red) can be generated by a string of Pauli $X$ operators (labelled in red) on qubits on vertical edges joining it to the top (or bottom) boundary.}
\label{fig:compact_mapping_syndrome}
\end{figure}

Note that after step 2, the combined state of the primary and auxiliary qubits satisfies
\begin{align}
s_i^p\otimes s_i^a \ket{\phi_{det}}_{p}\otimes\ket{\phi_{surf}}_{a}&=b_i\ket{\phi_{det}}_{p}\otimes\ket{\phi_{surf}}_{a}
\end{align}
for each generator $s_i=s_i^p\otimes s_i^a$ of $\mathcal{S}$, where the eigenvalue $b_i\in\{-1,1\}$ is the parity of the primary qubits acted on non-trivially by $s_i^p$, satisfying $s_i^p\ket{\phi_{det}}_{p}=b_i\ket{\phi_{det}}_{p}$.
We say that $b_i$ is the syndrome of generator $s_i$.

In step 3, we apply a circuit that instead ensures that we are in the $+1$-eigenspace of elements of $\mathcal{S}$.
This can be done by applying a Pauli operator $R$, with support only on the auxiliary qubits, that commutes with each generator $s_i$ if its syndrome $b_i$ is 1 and anti-commutes otherwise.
Such a Pauli operator can always be found for any assignment of each $b_i\in\{1,-1\}$, as shown in \Cref{fig:compact_mapping_syndrome}: for each stabiliser generator $s_i$, we can find a Pauli operator that we denote $V(s_i)$ which, acting only on the auxiliary qubits, anti-commutes with $s_i$ while commuting with all other generators (note that the choice of $V(s_i)$ is not unique).
Taking the product of operators $V(s_i)$ for all $s_i$ with syndrome $b_i=-1$, we obtain a single Pauli operator 
\begin{equation}\label{eq:compact_correction}
R=\prod_{i\in\{i: b_i=-1\}}V(s_i)
\end{equation}
that returns the state of our combined system to the $+1$-eigenspace of elements of $\mathcal{S}$, such that it satisfies \Cref{eq:determinant_stabilisers}.
Furthermore, since steps 2 and 3 have acted trivially on the primary qubits, \Cref{eq:vertex_op_eigenvalues} is still satisfied from step 1.
Therefore, a Slater determinant in the compact mapping can be encoded using the $O(L)$ depth unitary encoding circuit for the planar code as well as $O(1)$ layers of single qubit Clifford gates.
Note that the topologically protected logical qubit in the compact mapping is not used to store quantum information.
As a result, we can prepare any state in the codespace of the surface code in step 2, and it does not matter if the Pauli correction $R$ in step 3 acts non-trivially on the logical qubit.
The problem of finding a suitable correction $R$ in step 3 given the syndrome of each generator is essentially the same problem as decoding the XZZX surface code~\cite{wen2003quantum,bonilla2020xzzx} under the quantum erasure channel (and where every qubit is erased).
Therefore, any other suitable decoder could be used instead of using \Cref{eq:compact_correction}, such as the variant of minimum-weight perfect matching used in Ref.~\cite{bonilla2020xzzx}, or an adaptation of the peeling decoder~\cite{delfosse2020linear}.

The encoding step for the surface code could instead be done using stabiliser measurements. However, since it is not otherwise necessary to measure the stabilisers of the mapping, the additional complexity of using ancillas, mid-circuit measurements and real-time classical logic might make such a measurement-based approach more challenging to implement on either NISQ or fault-tolerant hardware than the simple $O(L)$ depth local unitary encoding circuit we present. Furthermore, the $O(L)$ complexity of our encoding circuit is likely negligible compared to the overall complexity of most quantum simulation algorithms within which it could be used.
Our encoding circuits for the surface code may also be useful for preparing states encoded in other fermion-to-qubit mappings. 
As an example, it has previously been observed that the Verstraete-Cirac transform also has a similar stabiliser structure to the surface code~\cite{verstraete2005mapping,steudtner2019quantum}.

\hspace{0pt}

\section{Discussion}

We have presented local unitary circuits for encoding an unknown state in the surface code that take time linear in the lattice size $L$. Our results demonstrate that the $\Omega(L)$ lower bound given by Bravyi \textit{et al.}~\cite{bravyi2006lieb} for this problem is tight, and reduces the resource requirements for experimentally realising topological quantum order and implementing some QEC protocols, especially using NISQ systems restricted to local unitary operations. We have provided a new technique to encode the planar code in $O(L)$ time, as well as showing how an $O(L)$ local unitary encoding circuit for the toric code can be found by enforcing locality in the non-local RG encoder. We unify these two approaches by demonstrating how local $O(L)$-depth circuits can be used to convert between the planar and toric code, and generalise our method to rectangular, rotated and 3D surface codes.

We also show that our unitary encoding circuit for the planar code can be used to encode a Slater determinant state in the compact mapping~\cite{derby2021compact}, which has a similar stabiliser structure to the surface code.
This encoding circuit is therefore a useful subroutine for the simulation of fermionic systems on quantum computers, and it may be that similar techniques can be used to encode fermionic states in the Verstraete-Cirac transform, which has a similar stabiliser structure~\cite{verstraete2005mapping}.

Using known local unitary mappings from one or more copies of the surface code, our results also imply the existence of optimal encoders for any 2D translationally invariant topological code, some 2D subsystem codes~\cite{yoshida2011classification, bombin2012universal}, as well as the 2D color code with and without boundaries~\cite{kubica2015unfolding}. As an explicit example, the subsystem surface code with three-qubit check operators can be encoded from the toric code using the four time step quantum circuit given in Ref.~\cite{bravyi2012subsystem}.

The circuits we have provided in this work are not fault-tolerant for use in error correction: a single qubit fault at the beginning of the circuit can lead to a logical error on the encoded qubit.
Nevertheless, since our circuits have a lower depth than local unitary circuits given in prior work, we expect our circuits also to be more resilient to circuit noise (for example, our circuits have fewer locations for an idle qubit error to occur).
Fault-tolerance of the encoding circuit itself is also not required when using it to prepare fermionic states or to study topological quantum order: for these applications, our circuits could be implemented using either physical qubits (on a NISQ device) or logical qubits on a fault-tolerant quantum computer.
It would be interesting to investigate if our circuits could be adapted to be made fault-tolerant, perhaps for the preparation of a known state (e.g.~logical $\ket{0}$ or $\ket{+}$).
Further work could also investigate optimal local unitary encoding circuits for surface codes based on different lattice geometries (such as the hexagonal lattice~\cite{fujii2012error}),  or for punctured~\cite{raussendorf2007fault, fowler2009high} or hyperbolic surface codes~\cite{breuckmann2016constructions}.

\begin{acknowledgments}
The authors would like to thank Mike Vasmer for informing us of the method for encoding stabiliser codes in Ref.~\cite{gottesman1997stabilizer}, as well as Charlie Derby and Joel Klassen for insightful discussions on fermion-to-qubit mappings. We are also grateful for helpful discussions with Austin Fowler and Benjamin Brown, and would like to thank Selwyn Simsek and Adam Callison for pointing out a formatting error in an earlier version of this manuscript. We thank Engineering and Physical Sciences Research Council (EPSRC) for funding this work.  Dan Browne and Simon Burton were funded by EPSRC grant EP/R043647/1 and the remaining authors by EPSRC grant number EP/L015242/1. In addition, Farhan Hanif and James Dborin gratefully acknowledge funding from University College London.
\end{acknowledgments}

\textit{Note added}: After the first preprint of this article, Ref.~\cite{satzinger2021realizing} introduced an alternative $O(L)$ unitary encoding circuit for the rotated surface code, using it to experimentally realise topological quantum order.

\bibliographystyle{plainnat}
\bibliography{bibliography}

\appendix

\section{Procedure for Encoding a Stabiliser Code}\label{app:gottesman}
\subsection{Review of the General Method}
\normalsize
In this section we review the general method for constructing an encoding circuit for arbitrary stabiliser codes given in \cite{gottesman1997stabilizer, Cleve_1997}, and show how it can be used to find an encoding circuit for an $L=2$ toric code as an example.
We present the method here for completeness, giving the procedure in full and in the simplified case for which the code is CSS. 

From a set of check operators one can produce a corresponding bimatrix
\[
M\coloneqq
\begin{pmatrix}[c|c]
  L & R\\
\end{pmatrix}
\]
Rows and columns represent check operators and qubits respectively. $L_{ij} =1$ indicates that check operator $i$ applies $X$ to qubit $j$ as opposed to the identity, similarly for the right hand side $R_{ij} = 1$ implies check operator $i$ applies $Z$ to qubit $j$. If both $L_{ij} = 1$ and $R_{ij} = 1$, then check operator $i$ applies $Y$ on qubit $j$.

A CSS code has check operators $P_n \in \{I,X\}^{\otimes n } \cup \{I,Z\}^{\otimes n }$, its corresponding bimatrix takes the form,
\[
\begin{pmatrix}[c|c]
  A & 0\\
  0 & B
\end{pmatrix}
\]

$A$ and $B$ have full row rank since they each represent an independent subset of the check operators. Labelling the rank of $A$ as $r$, the rank of $B$ is $n-k-r$.

Via row addition, row swaps and column swaps, the left and right matrices of this simplified form can be taken to standard form \cite{gottesman1997stabilizer} without changing the stabiliser group of the code. The standard form of the bimatrix is then

\[
\begin{pmatrix}[ccc|ccc]
  I & A_1 & A_2 & B & C_1 & C_2\\
  0 & 0 & 0 & D & I & E
\end{pmatrix}
\]
Where $I$,$A_1$,$A_2$ and $D,I,E$ have $(r)$, $(n-k-r)$, and $(k)$ columns respectively. We may also represent the set of logical $X$ operators as a bimatrix with each row representing the logical $X$ for a particular encoded qubit,

\[ \bar{X} = 
\begin{pmatrix}[ccc|ccc]
  U_1 & U_2 & U_3 & V_1 & V_2 & V_3\\
\end{pmatrix}
\]

It is shown in \cite{gottesman1997stabilizer} that the logical $\bar{X}$ operator can be taken to the form 

\[ \bar{X} = 
\begin{pmatrix}[ccc|ccc]
  0 & U_2 & I & V_1 & 0 & 0\\
\end{pmatrix}
\]

In the CSS case the check operator bimatrix reduces to
\[
\begin{pmatrix}[ccc|ccc]
  I & A_1 & A_2 & 0 & 0 & 0\\
  0 & 0 & 0 & D & I & E
\end{pmatrix}
\]

and the logical $X$ bimatrix to
\[ \bar{X} = 
\begin{pmatrix}[ccc|ccc]
  0 & E^T & I & 0 & 0 & 0\\
\end{pmatrix}
\]

To produce a circuit which can encode state $\ket{c_1 \dots c_k}$ for any values of the $c_i$ one should find a circuit which applies logical operators $\bar{X}_1^{c_1} \dots \bar{X}_{k}^{c_k}$ to the encoded $\ket{0}$ state $\ket{\bar{0}} \equiv \sum_{S \in \mathcal{S}}S \ket{0 \dots 0 }$.
Let $F_{c}$ be the operator corresponding to row $c$ of bimatrix $F$. 
We denote by $F_{c(m)}$ the operator corresponding to $F_c$, with the operator on the $m^{th}$ qubit replaced with identity, and then controlled by the $m^{th}$ qubit.

Since
\[ \bar{X}_1^{c_1} \dots \bar{X}_k^{c_k}  \sum_{S \in \mathcal{S}}S \ket{0 \dots 0 }  =   \sum_{S \in \mathcal{S}}S \bar{X}_1^{c_1} \dots \bar{X}_k^{c_k} \ket{0 \dots 0 } \]
the application of the $X$ gates can be considered before applying the sum of stabiliser operations. Due to the $I$ in the form of $\bar{X}$,
\[\bar{X}_{k(n)} \ket{0_1 \dots 0_{n-k}}\ket{0_{n-k+1} \dots 0_{n-1} c_k} = \bar{X}^{c_k}_{k}\ket{0_1 \dots 0_n}.\] we see that independently of $\ket{c_1 \dots c_k}$ we can implement 

\begin{align*}
     \bar{X}_{1(n-k)} \dots \bar{X}_{k(n)} & \ket{0_1 \dots 0_{n-k}}  \ket{c_1 \dots c_k} \\
    & =   \bar{X}^{c_1}_{1} \dots \bar{X}^{c_k}_{k} \ket{0_1 \dots 0_n} \\
    & \equiv \ket{0_1 \dots 0_r}\ket{Xc}
\end{align*}
where in the last line it is emphasised that since $U_1 = 0$, $X_{i(j)}$ acts trivially on the first $r$ qubits. Next to consider is $\sum_{S \in \mathcal{S}}S =  (I + M_{n-k}) \dots (I + M_r) \dots (I + M_1)$.

We denote the right matrix of bimatrix $M$ as $R$. In standard form $M_i$ always performs $X$ on qubit $i$ and it performs $Z$ on qubit $i$ when $R_{ii} = 1$, giving
\begin{align*}
    M_{i}\ket{0 \dots 0_{i} \dots 0} = Z_{i}^{R_{ii}}M_{i(i)}\ket{0 \dots 1_{i} \dots 0}
\end{align*}
and so
\begin{align*}
    (I+M_{i})\ket{0 \dots 0} & = \ket{0 \dots 0_{i} \dots 0} + M_{i}\ket{0 \dots 0_{i} \dots 0} \\ & = Z_{i}^{R_{ii}}M_{i(i)}H_{i}\ket{0 \dots 0}
\end{align*}
or generally
\begin{align*}
     & \prod_{i = 1}^{r}(I + M_{i}) (\ket{0_1 \dots 0_r}\ket{Xc}) \\
    =  & \prod_{i = 1}^{r} Z_{i}^{R_{ii}}M_{i(i)}H_{i} (\ket{0_1 \dots 0_r}\ket{Xc})
\end{align*}
The remaining products
\begin{equation}
     \prod_{i = r+1}^{n-k}(I + M_{i})
\end{equation}
can be ignored since they consist only of $\sigma_{z}$ operations and may be commuted to the front to act on $\ket{0}$ states.

Given initially some $k$ qubits we wish to encode, and some additional $n-k$ auxiliary qubits, initialised in $\ket{0}$, a choice of generators for the stabiliser group is 

\[
\begin{pmatrix}[cccccccc|cccccccc]
  0 & 0 & 0 & 0 & 0 & 0 & 0 & 0 & 1 & 0 & 0 & 0 & 0 & 0 & 0 & 0\\
  0 & 0 & 0 & 0 & 0 & 0 & 0 & 0 & 0 & 1 & 0 & 0 & 0 & 0 & 0 & 0\\
  0 & 0 & 0 & 0 & 0 & 0 & 0 & 0 & 0 & 0 & 1 & 0 & 0 & 0 & 0 & 0\\
  0 & 0 & 0 & 0 & 0 & 0 & 0 & 0 & 0 & 0 & 0 & 1 & 0 & 0 & 0 & 0\\
  0 & 0 & 0 & 0 & 0 & 0 & 0 & 0 & 0 & 0 & 0 & 0 & 1 & 0 & 0 & 0\\
  0 & 0 & 0 & 0 & 0 & 0 & 0 & 0 & 0 & 0 & 0 & 0 & 0 & 1 & 0 & 0\\
\end{pmatrix}
\]

The general circuit which transforms the initial generator set to the standard form bimatrix is given by,
\[
\prod_{i = 1}^{r}Z_{i}^{R_{ii}}M_{i(i)}H_{i} \prod_{j=1}^{k}\bar{X}_{j(n-k+j)}  
\]
For CSS codes this reduces to 

\[
\prod_{i = 1}^{r}M_{i(i)}H_{i} \prod_{j=1}^{k}\bar{X}_{j(n-k+j)}  
\]

In the simplified case all gates are either initial $H$ gates or $CNOT$'s. We may write the circuit in two stages, performing first the $H$ gates and controlled $\bar{X}$ gates.

\begin{equation*}
\Qcircuit @C=1em @R=.7em {
& \gate{H} & \qw & \qw & \qw & \qw & \qw & \qw & \qw\\
& \cdots &  &  & & &  &  & \\
& \gate{H} & \qw & \qw & \qw & \qw & \qw & \qw & \qw\\
& \qw & \multigate{2}{{E}_{1(1)}} & \qw & \qw & \qw & \multigate{2}{{E}_{k(k)}} & \qw & \qw \\
& \cdots & \nghost{{E}_{1(1)}} & \cdots & & \cdots & \nghost{{E}_{k(k)}} & \cdots & \\
& \qw & \ghost{{E}_{1(1)}} & \qw & \qw & \qw & \ghost{{E}_{k(k)}} & \qw & \qw\\
& \qw & \qw & \qw & \cdots&  & \ctrl{-1} & \qw & \qw\\
& \cdots &  &  &  &  &  &  & \\
& \qw & \ctrl{-3} & \qw & \cdots &  & \qw & \qw & \qw \\
}
\end{equation*}
and in stage 2 the controlled $X$ gates.

\begin{equation*}
\Qcircuit @C=1em @R=.7em {
& \qw & \qw & \ctrl{3} & \qw & \cdots &  & \qw & \qw\\
& \cdots &  &   & & &  &  & \\
& \qw & \qw & \qw & \qw & \cdots &  & \ctrl{1} & \qw\\
& \qw & \qw & \multigate{5}{M_{1(1)}} & \qw & \cdots &  & \multigate{5}{M_{r(r)}} & \qw \\
& \cdots &  & \nghost{M_{1(1)}} &  & \cdots &  &  & \\
& \qw & \qw & \ghost{M_{1(1)}} & \qw & \cdots &  & \ghost{M_{r(r)}} & \qw\\
& \qw & \qw & \ghost{M_{1(1)}} & \qw & \cdots &  & \ghost{M_{r(r)}} & \qw\\
& \cdots &  & \nghost{M_{1(1)}} &  &  &  & \nghost{M_{r(r)}} & \\
& \qw & \qw & \ghost{M_{1(1)}} & \qw & \cdots &  & \ghost{M_{r(r)}} & \qw \\
}
\end{equation*}

In the general case stage 1 is identical but stage 2 takes the form 

\begin{equation*}
\Qcircuit @C=1em @R=.7em {
& \qw & \multigate{2}{\Omega_{z}} & \ctrl{3} & \qw & \cdots &  & \multigate{1}{M_{r(r)}} & \\
& \cdots & \nghost{\sigma_{z}}  & \qw  & \qw & &  & \ghost{M_{r(r)}} &  \\
& \qw & \ghost{\sigma_{z}} & \qw & \qw & \cdots &  & \ctrl{1} \qwx & \qw\\
& \qw & \qw & \multigate{5}{M_{1(1)}} & \qw & \cdots &  & \multigate{5}{M_{r(r)}} & \qw \\
& \cdots &  & \nghost{M_{1(1)}} &  & \cdots &  &  & \\
& \qw & \qw & \ghost{M_{1(1)}} & \qw & \cdots &  & \ghost{M_{r(r)}} & \qw\\
& \qw & \qw & \ghost{M_{1(1)}} & \qw & \cdots &  & \ghost{M_{r(r)}} & \qw\\
& \cdots &  & \nghost{M_{1(1)}} &  &  &  & \nghost{M_{r(r)}} & \\
& \qw & \qw & \ghost{M_{1(1)}} & \qw & \cdots &  & \ghost{M_{r(r)}} & \qw \\
}
\end{equation*}

Where $\Omega_{z}$ consists of $Z$ operations on some of the first $r$ qubits and each $M_{i(i)}$ consists of controlled $Z$ gates on some of the first $r$ qubits and controlled Pauli gates on some of the following $n-r$ qubits. In the case of the $L=2$ toric code, with qubits labelled left to right and top to bottom the bimatrix is 

\[
\begin{pmatrix}[cccccccc|cccccccc]
  1 & 1 & 1 & 0 & 0 & 0 & 1 & 0 & 0 & 0 & 0 & 0 & 0 & 0 & 0 & 0\\
  1 & 1 & 0 & 1 & 0 & 0 & 0 & 1 & 0 & 0 & 0 & 0 & 0 & 0 & 0 & 0\\
  0 & 0 & 1 & 0 & 1 & 1 & 1 & 0 & 0 & 0 & 0 & 0 & 0 & 0 & 0 & 0\\
  0 & 0 & 0 & 0 & 0 & 0 & 0 & 0 & 1 & 0 & 1 & 1 & 1 & 0 & 0 & 0\\
  0 & 0 & 0 & 0 & 0 & 0 & 0 & 0 & 0 & 1 & 1 & 1 & 0 & 1 & 0 & 0\\
  0 & 0 & 0 & 0 & 0 & 0 & 0 & 0 & 1 & 0 & 0 & 0 & 1 & 0 & 1 & 1\\
\end{pmatrix}
\]

The standard form of this bimatrix is

\[
\begin{pmatrix}[cccccccc|cccccccc]
  1 & 0 & 0 & 1 & 0 & 1 & 0 & 1 & 0 & 0 & 0 & 0 & 0 & 0 & 0 & 0\\
  0 & 1 & 0 & 0 & 1 & 1 & 0 & 1 & 0 & 0 & 0 & 0 & 0 & 0 & 0 & 0\\
  0 & 0 & 1 & 0 & 0 & 1 & 1 & 1 & 0 & 0 & 0 & 0 & 0 & 0 & 0 & 0\\
  0 & 0 & 0 & 0 & 0 & 0 & 0 & 0 & 0 & 1 & 1 & 1 & 0 & 0 & 0 & 1\\
  0 & 0 & 0 & 0 & 0 & 0 & 0 & 0 & 0 & 1 & 1 & 0 & 1 & 0 & 1 & 0\\
  0 & 0 & 0 & 0 & 0 & 0 & 0 & 0 & 1 & 1 & 1 & 0 & 0 & 1 & 0 & 0\\
\end{pmatrix}
\]

The circuit which encodes the above stabiliser set is
\begin{equation*}
\Qcircuit @C=1em @R=.7em {
& \gate{H} & \qw & \ctrl{3}& \ctrl{5} & \ctrl{7} & \qw &\qw &\qw &\qw &\qw &\qw \\ 
& \gate{H} & \qw & \qw & \qw & \qw &\ctrl{3} &\ctrl{4}&\ctrl{6} &\qw &\qw&\qw \\ 
& \gate{H} & \qw & \qw & \qw& \qw & \qw &\qw &\qw &\ctrl{3} &\ctrl{4}&\ctrl{5} \\ 
& \qw & \targ & \targ & \qw & \qw &\qw &\qw &\qw &\qw &\qw&\qw \\
& \targ & \qw & \qw & \qw &\qw &\targ &\qw &\qw &\qw &\qw&\qw \\ 
& \qw & \qw & \qw & \targ &\qw &\qw &\targ &\qw &\targ &\qw&\qw \\
& \ctrl{-2} & \qw & \qw & \qw &\qw &\qw &\qw &\qw &\qw &\targ&\qw \\ 
& \qw & \ctrl{-4} & \qw & \qw &\targ &\qw &\qw &\targ &\qw &\qw&\targ } 
\end{equation*}

It is important to have kept track of which column represents which qubit since column swaps are performed in bringing the matrix to standard form. Taking this into account gives the $L=2$ circuit on the toric architecture.

\subsection{Depth of the General Method}

Any stabiliser circuit has an equivalent skeleton circuit~\cite{Maslov_2007} (a circuit containing only generic two-qubit gates, with single-qubit gates ignored) which after routing on a surface architecture will have at worst $O(n)$ depth. The output of the general method for encoding a stabiliser code in fact already splits into layers of skeleton circuits. Stage 2 of the method applied to a CSS code has at worst $r(n-r)$ controlled Pauli gates $CP_{ij}$ with $i$,$j$ in $\{1 \dots r\}$ and $\{r+1 \dots n\}$ respectively, $CP_{ij}$ is implemented before $CP_{i'j'}$ so long as $i<i'$. Stage 2 then takes the form of a skeleton circuit and as such the number of timesteps needed is $O(n)$ for surface or linear nearest neighbour architectures. Stage 1 has at most $k(n-k-r)$ gates and also takes the form of a skeleton circuit. In the worst case scenario stage 2 includes, in addition to the $CP$ gates, controlled Z gates $CZ$ with targets on the first $r$ qubits. As noted in errata for \cite{gottesman1997stabilizer}, $i>j$ for any of the additional $CZ_{ij}$ in stage 2. All $CZ_{ij}$ can then be commuted to timesteps following all $CP$ gates since each $CP$ in a timestep following $CZ_{ij}$ takes the form $CP_{mn}$ with $n>m>i>j$. The circuit then splits into a layer of $CP$ gates and a layer of $CZ$ gates, each of which is a skeleton circuit, and so can be implemented in $O(n)$ timesteps on surface and linear nearest neighbour architectures.

\section{Additional planar encoding circuits}\label{app:additional_planar_encoders}
\subsection{Planar base cases and rectangular code}
\begin{figure}
    \centering
    \subfloat[\label{subfig:L2_planar}]{
        \includegraphics[width=0.35\columnwidth, trim={0 0 5.5cm 0}]{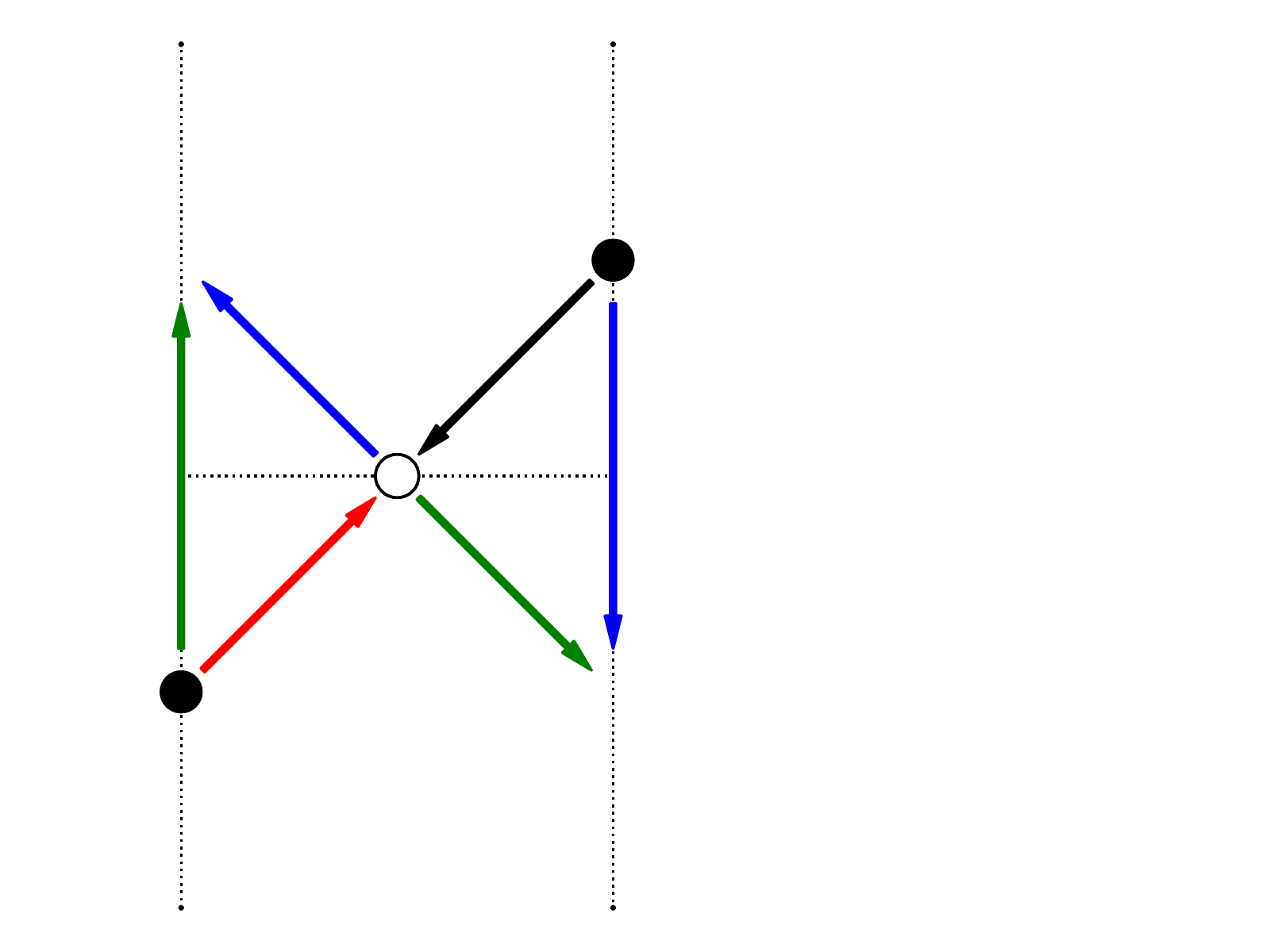}
     }
     \subfloat[\label{subfig:L3_planar}]{
        \includegraphics[width=0.4\columnwidth, trim={0 0 4cm 0}]{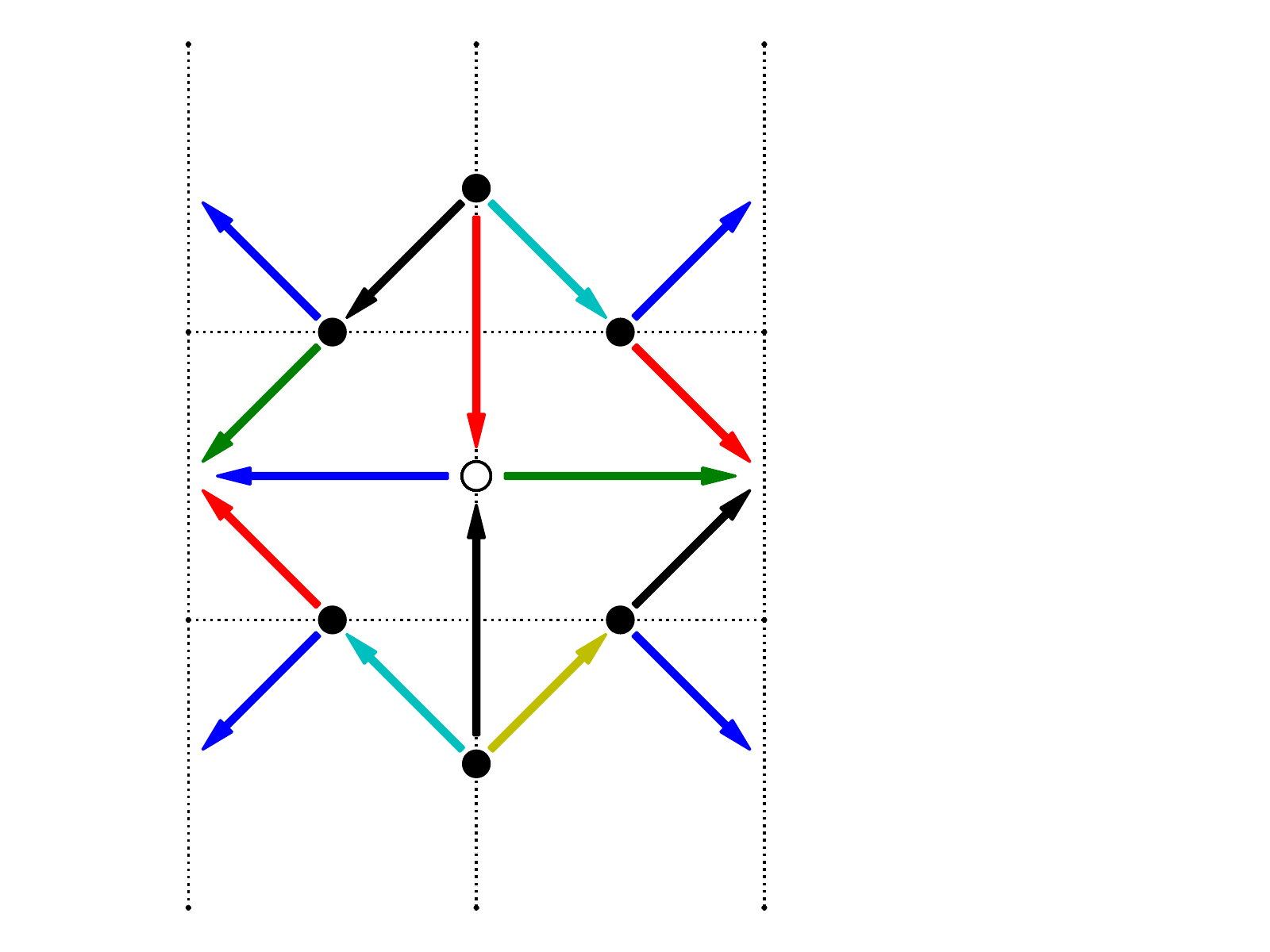}
     }
     \\
     \subfloat[\label{subfig:L4_planar}]{
        \includegraphics[trim={0 0 3cm 0}, width=0.5\columnwidth]{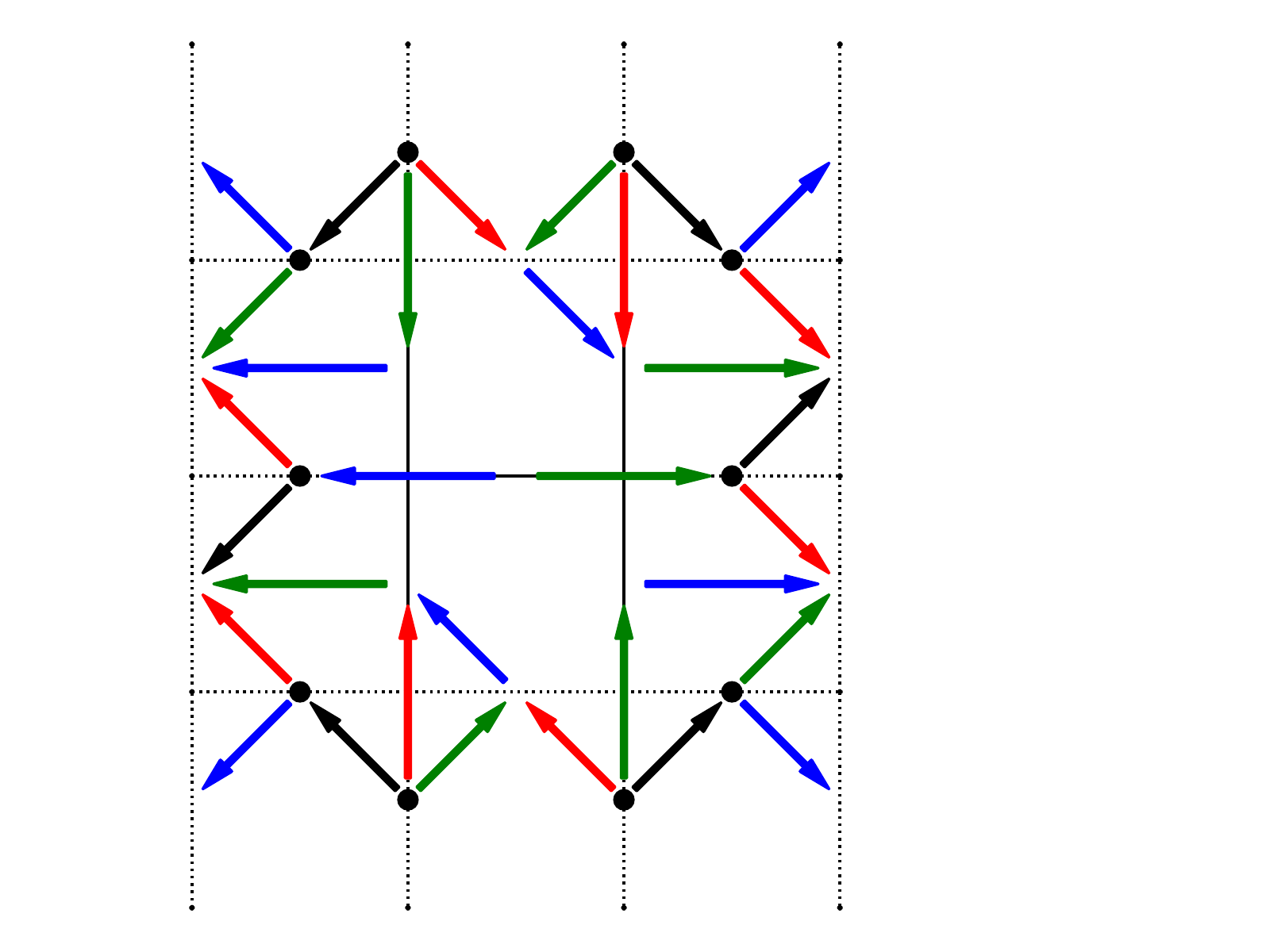}
     }
    \caption{Encoding circuits for the L=2, L=3 and L=4 planar codes. Each edge corresponds to a qubit, each arrow denotes a CNOT gate pointing from control to target, and each filled black circle denotes a Hadamard gate applied at the beginning of the circuit. The colour of each CNOT gate corresponds to the time step it is implemented in, with blue, green, red, black, cyan and yellow CNOT gates corresponding to the first, second, third, fourth, fifth and sixth time steps respectively. The hollow circle in each of (a) and (b) denotes the initial unencoded qubit. The circuit in (c) encodes an L=4 planar code from an L=2 planar code, with solid edges denoting qubits initially encoded in the L=2 code.}
    \label{fig:planar_base_cases}
\end{figure}

\begin{figure}
    \centering
    \subfloat[]{
    \includegraphics[width=0.58\columnwidth,trim={0 0 0 0}]{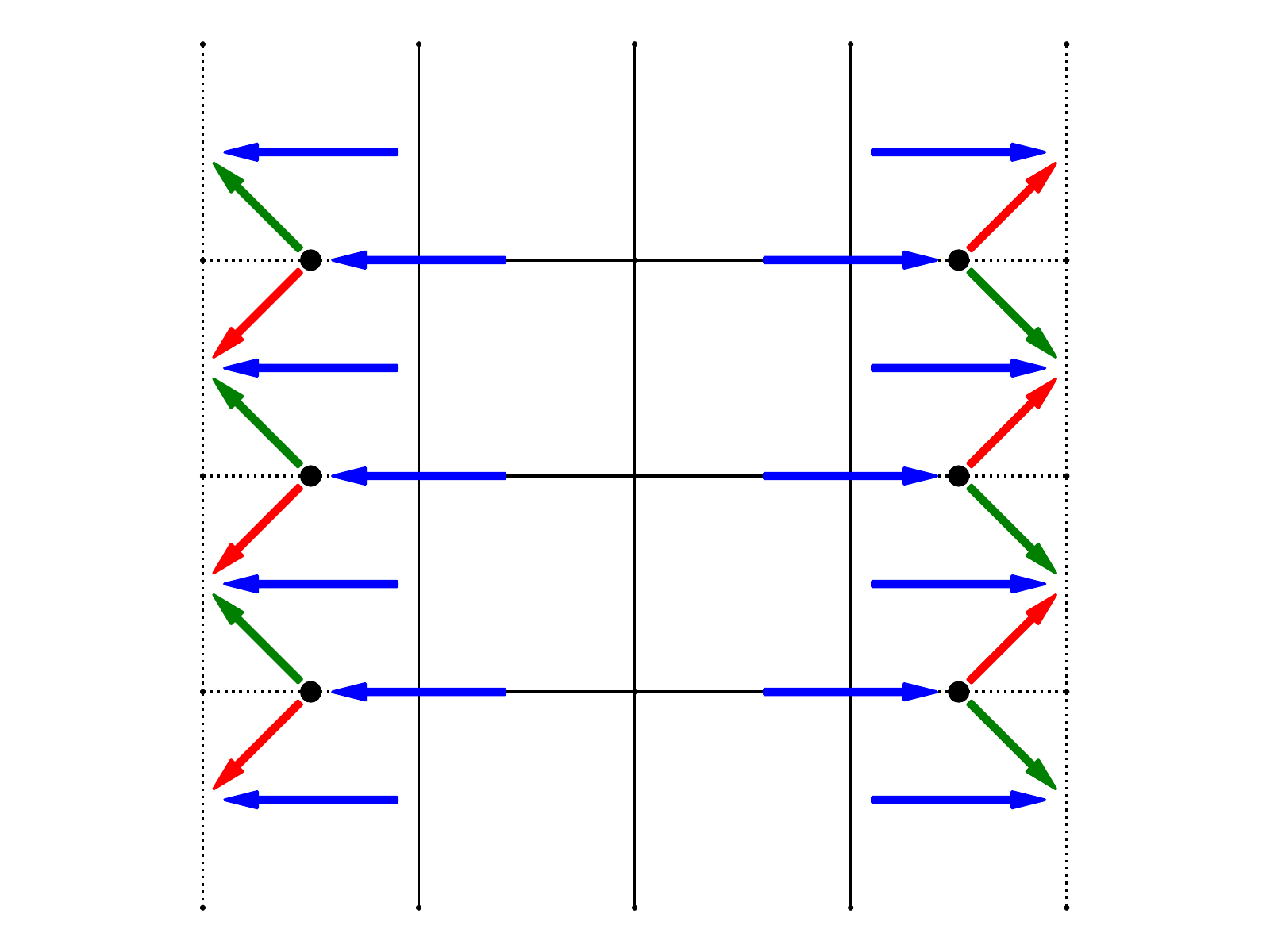}
    }
    \subfloat[]{
    \includegraphics[width=0.3\columnwidth,trim={2cm 0 6.5cm 0}]{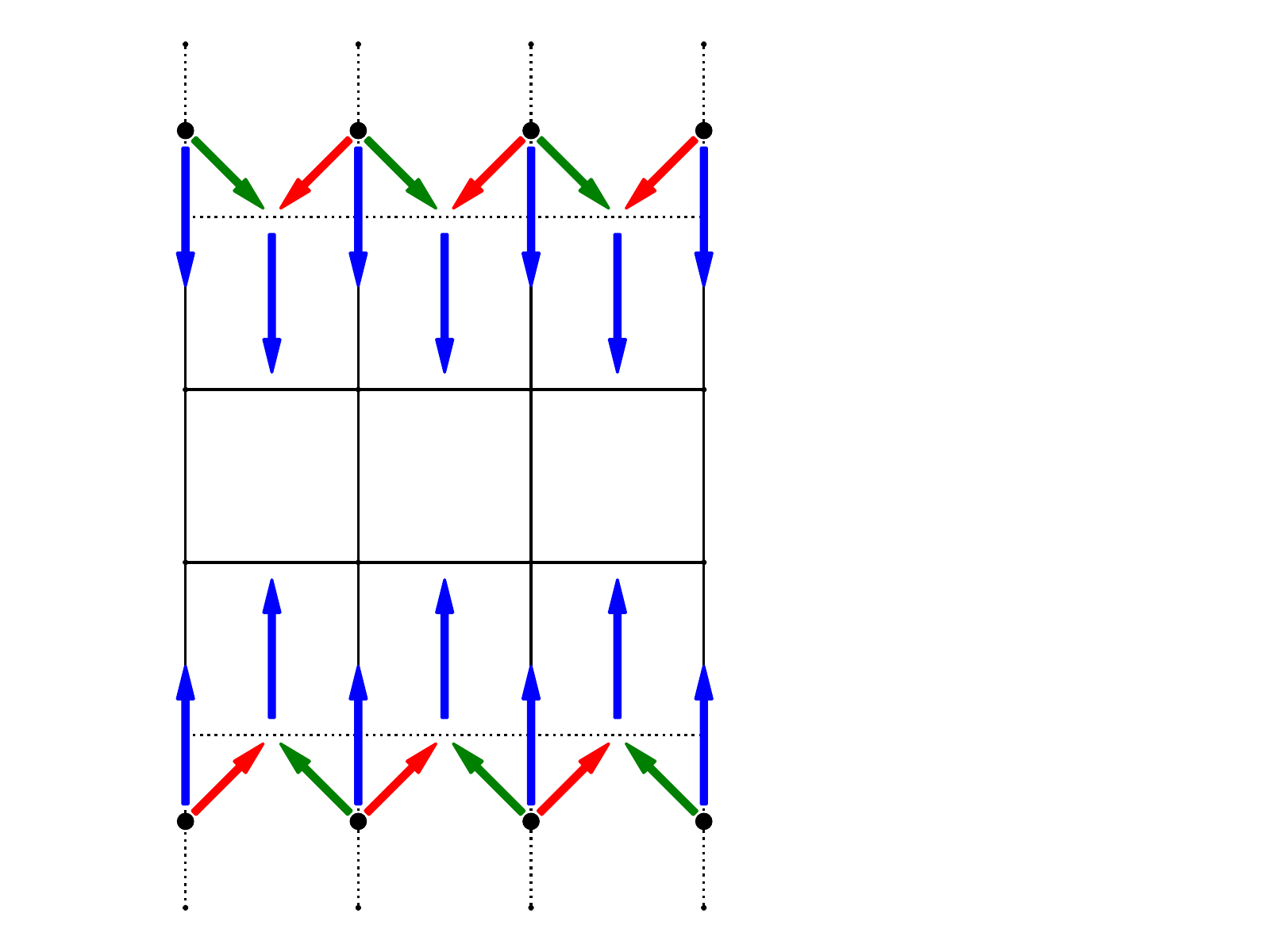}
    }
    \caption{(a) Circuit to increase the width of a planar code by two. (b) Circuit to increase the height of a planar code by two. Notation is the same as in \Cref{fig:planar_base_cases}.}
    \label{fig:rectangular_code}
\end{figure}

In \Cref{fig:planar_base_cases} we provide encoding circuits for the $L=2$, $L=3$ and $L=4$ planar codes, requiring 4, 6 and 8 time steps respectively. These encoding circuits are used as base cases for the planar encoding circuits described in \Cref{sec:planar_encoding}. In \Cref{fig:rectangular_code} we provide encoding circuits that either increase the width or height of a planar code by two, using three time steps.

\subsection{Rotated Surface Code}\label{app:rotated_code}

\begin{figure}
    \centering
    \includegraphics[width=0.5\textwidth]{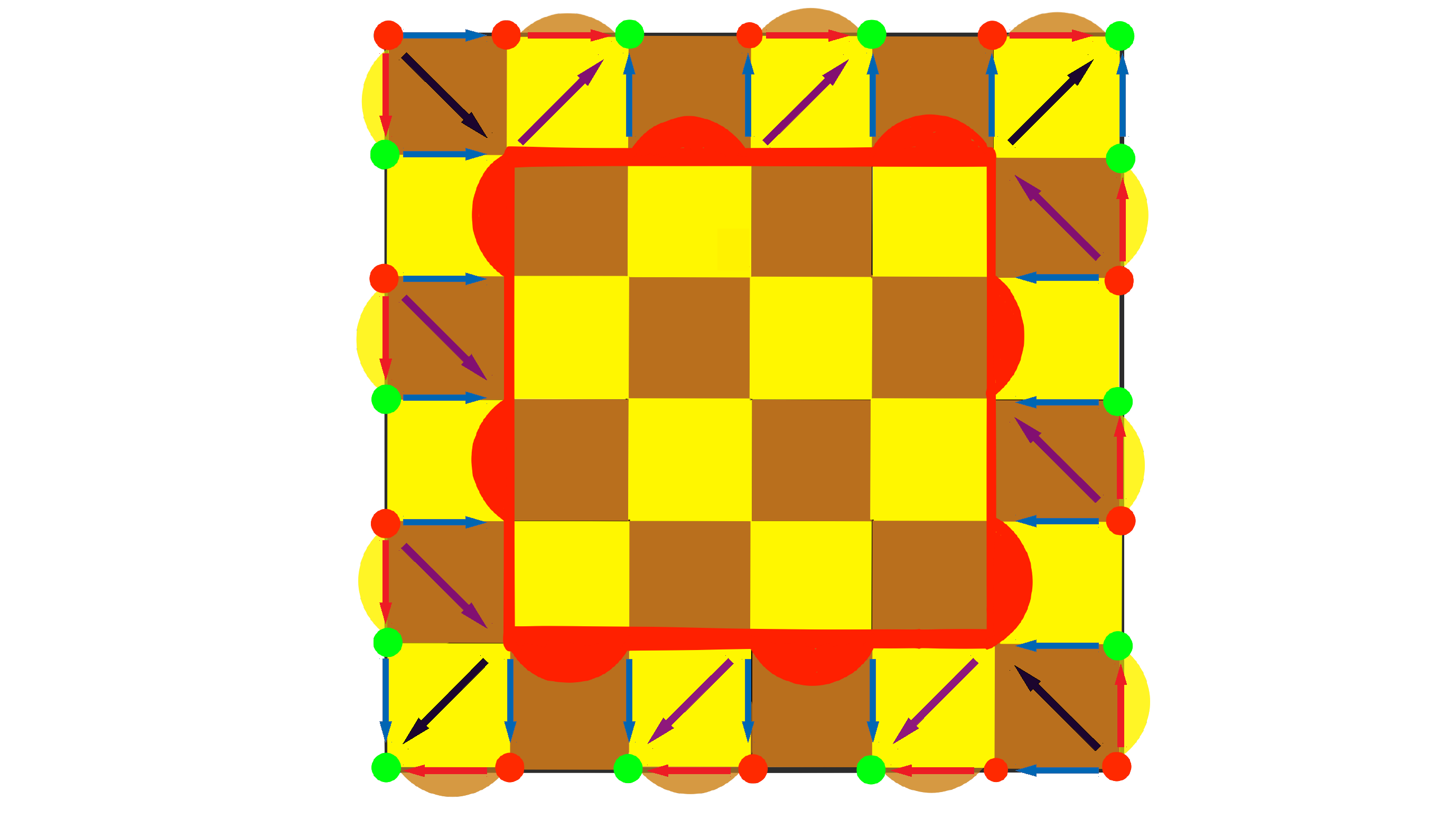}
    \caption{Encoding circuit for the $L=7$ rotated code from an $L=5$ rotated surface code (shown as a red outline). The colour of each arrow denotes the time step the gate is applied in. The gates are applied in the order: blue, red, black, purple. 
    The additional qubits are initialised in the $\ket{+}$ (red) or $\ket{0}$ (green) state. 
    The yellow squares denote a $Z$ stabiliser on the four corner qubits, and the brown squares represent an $X$ operator on the four corner qubits.
    The rotated code has additional stabilizers between states on along the edges. In the $L=5$ code these are shown as a red arch (with $Z$ and $X$ stabilisers on the vertical and horizontal edges respectively), and the yellow and brown arches in the $L=7$ code edge are Z and X stabilizers between the two edge qubits.}
    \label{fig:rotated_code}
\end{figure}

In \Cref{fig:rotated_code} we demonstrate a circuit that encodes an $L=7$ rotated surface code from a distance $L=5$ rotated code. For a given distance $L$, the rotated surface code uses fewer physical qubits than the standard surface code to encode a logical qubit~\cite{bombin2007optimal}. Considering a standard square lattice with qubits along the edges, a rotated code can be produced by removing qubits along the corners of the lattice boundary, leaving a diamond of qubits from the centre of the original lattice. The diagram in \Cref{fig:rotated_code} shows the resultant code, rotated $45^{\circ}$ compared to the original planar code, and with each qubit now denoted by a vertex rather than an edge. For a distance $L$ code the rotated surface code requires $L^2$ qubits compared to $L^2 + (L-1)^2$ for the planar code.

The encoding circuit in \Cref{fig:rotated_code} takes 4 steps to grow a rotated code from a distance $L=5$ to $L=7$. This is a fixed cost for any distance $L$ to $L+2$. To produce a distance $L=2m$ code this circuit would be applied repeatedly $m+O(1)$ times to an $L=2$ or $L=3$ base case, requiring a circuit of total depth $2L + O(1)$. 
The circuit in \Cref{fig:rotated_code} can be verified by using \Cref{eq:cnotstabilisers} to see that a set of generators for the $L=5$ rotated code (along with the single qubit $Z$ and $X$ stabilisers of the ancillas) is mapped to a set of generators of the $L=7$ rotated code, as well as seeing that the $X$ and $Z$ logicals of the $L=5$ code map to the $X$ and $Z$ logicals of the $L=7$ rotated code.

\section{Renormalisation Group encoder}\label{app:renormalisation_group}

\subsection{Toric Code Encoder}\label{basetoric}
Applying the Gottesman encoder to the toric code, as shown in  Appendix~\ref{app:gottesman}, and then enforcing locality using SWAP gates, gives the following encoding circuit for the $L=2$ toric code that requires 10 time steps:
\begin{equation*}
\Qcircuit @C=0.8em @R=.7em {
\lstick{\ket{0}}& \gate{H} & \qw & \qw& \ctrl{1} & \qw & \ctrl{4} &\ctrl{3} &\targ &\ctrl{3} &\qw &\ctrl{3}&\targ&\ctrl{3}&\qw \\ 
\lstick{\ket{0}}& \qw & \qw & \targ & \targ & \qw &\qw &\qw&\qw &\qw &\qw&\qw&\qw&\qw&\qw \\ 
\lstick{\ket{\psi_0}}& \qw & \ctrl{1} & \qw & \qw& \targ & \qw &\qw &\qw &\qw &\qw&\qw&\qw&\qw&\qw \\ 
\lstick{\ket{0}}& \qw & \targ & \qw & \targ & \qw &\qw &\targ &\ctrl{-3} &\targ &\ctrl{2}&\targ&\ctrl{-3}&\targ&\qw \\
\lstick{\ket{0}}& \qw & \qw & \qw & \qw &\qw &\targ &\qw &\qw &\qw &\qw&\targ&\targ&\qw&\qw \\ 
\lstick{\ket{\psi_1}}& \qw & \qw & \ctrl{-4} & \qw &\qw &\targ &\qw &\qw &\qw &\targ&\qw&\qw&\targ&\qw \\
\lstick{\ket{0}}& \gate{H} & \qw & \qw & \qw &\ctrl{-4} &\qw &\qw &\qw &\qw &\qw&\qw&\ctrl{-2}&\ctrl{-1}&\qw\\ 
\lstick{\ket{0}}& \gate{H} & \qw & \qw & \ctrl{-4} &\qw &\ctrl{-2} &\qw &\qw &\qw &\qw&\ctrl{-3}&\qw&\qw&\qw }
\end{equation*}

where the qubits are numbered $0\ldots 7$ from top to bottom. This circuit encodes the initial unknown qubit states $\ket{\psi_0}$ and $\ket{\psi_1}$ into logical states $\ket{\bar{\psi_0}}$ and $\ket{\bar{\psi_1}}$ of an $L=2$ toric code with stabiliser group generators $X_0X_1X_2X_6$, $X_0X_1X_3X_7$, $X_2X_4X_5X_6$, $Z_0Z_2Z_3Z_4$, $Z_1Z_2Z_3Z_5$ and $Z_0Z_4Z_6Z_7$.

\begin{figure}
    \centering
    \includegraphics[trim={0 0 0 0}, width=0.85\columnwidth]{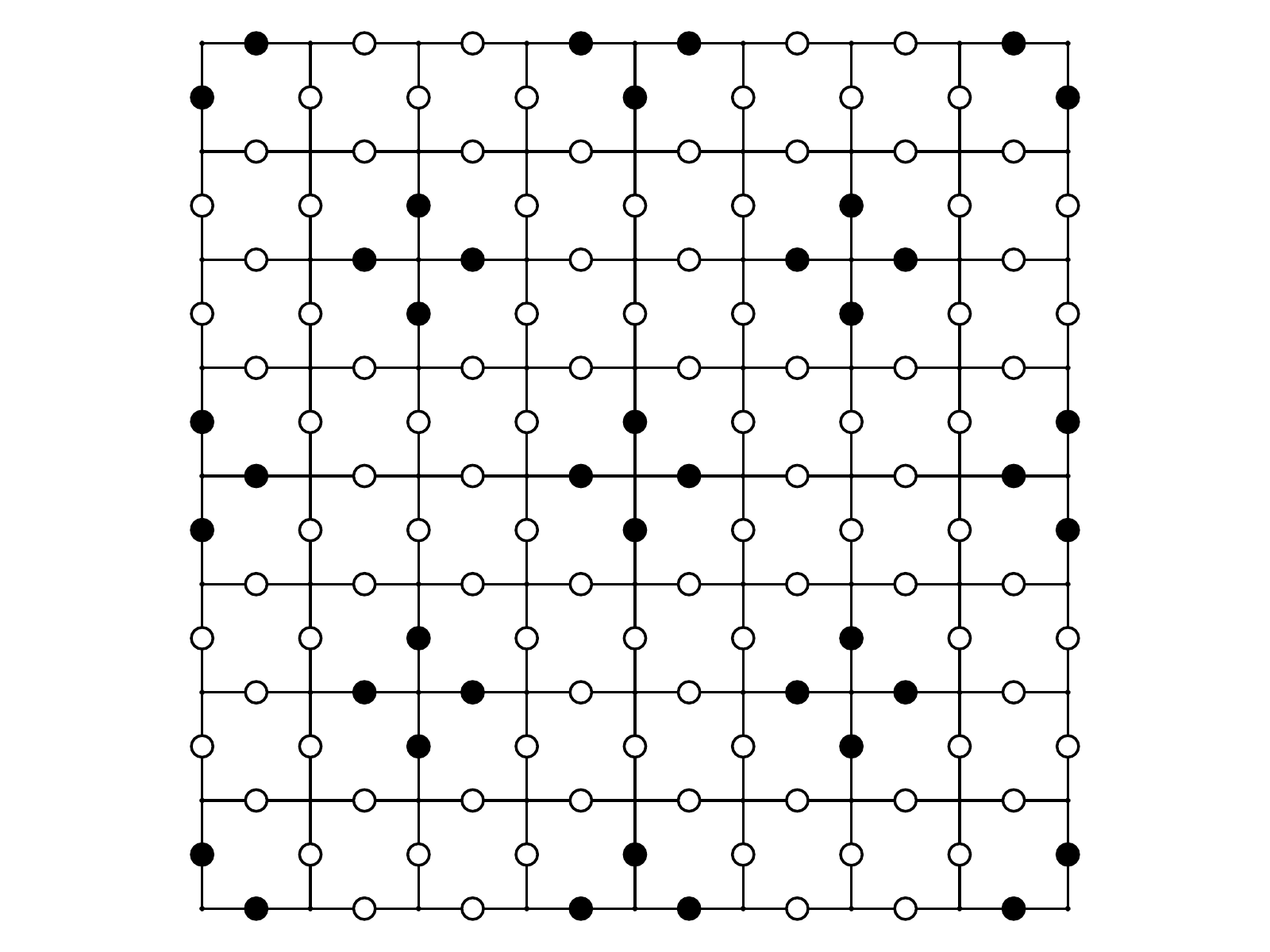}
    \caption{Initial outwards spreading of qubits in a distance 4 toric code to prepare for the encoding of a distance 8 toric code. Solid black and unfilled nodes represent the routed qubits of the distance 4 code, and the ancillae respectively. One then executes the subroutine of \Cref{fig:L2_to_L4} in each of the four 4x4 quadrants. This procedure generalises inductively for any targeted distance $2^k$ toric code.}
    \label{fig:L8_renorm}
\end{figure}

Equipped with an $L=2$ base code emulated as the central core of a $4\times 4$ planar grid, where the surrounding qubits are initially decoupled $+1$ $Z$ eigenstates, one can apply the local routing methods of Appendix \ref{routing} to obtain the initial configuration as depicted in \Cref{fig:L8_renorm}. The ancillae qubits are then initialised as $\ket{0}$ or $\ket{+}$ eigenstates as depicted in \Cref{fig:L2_to_L4}(a) by means of Hadamard operations where necessary, before the circuit is implemented through the sequence of CNOT gates as depicted in \Cref{fig:L2_to_L4}(a)-(c). 

By recursive application of \Cref{eq:cnotstabilisers}, it is seen that the circuit forms the stabiliser structure of an $L=4$ toric code on the planar architecture. Proceeding inductively, one can exploit the symmetry of a distance $L=2^k$ toric code to embed it in the centre of a $2L\times 2L$ planar grid, ``spread-out'' the core qubits in time linear in the distance, and ultimately perform the $L=2 \mapsto L=4$ circuit on each $4\times 4$ squarely-tesselated sub-grid.

\subsection{Routing circuits for enforcing locality}\label{routing}

To enforce locality in the Renormalisation Group encoder, which encodes a distance $L$ toric code, one can use SWAP gates to ``spread out'' the qubits between iteration $k$ and $k+1$, such that all of the $O(1)$ time steps in iteration $k+1$ are \textit{almost local} on a $2^{k+1}\times 2^{k+1}$ region of the $L\times L$ torus. By \textit{almost local}, we mean that the time step would be local if the $2^{k+1}\times 2^{k+1}$ region had periodic boundary conditions. Since at each iteration (until the final one) we use a region that is a subset of the torus, we in fact have a planar architecture (no periodic boundaries), and so it is not possible to simultaneously enforce locality in all of the $O(1)$ time steps in an iteration $k<\log L-2$ of the RG encoder, which are collectively local on a toric architecture. Thus it is necessary to emulate a toric architecture on a planar one. In a time step in iteration $k$, this can be achieved by using $3(2^{k}-1)$ time steps to move the top and bottom boundaries together (using SWAP gates) before applying any necessary gates which are now local (where the factor of three comes from the decomposition of a SWAP gate into 3 CNOT gates). Then $3(2^{k}-1)$ time steps are required to move the boundaries back to their original positions. The identical procedure can be applied simultaneously to the left and right boundaries. Thus there is an overhead of $3(2^{k+1}-2)$ to emulate a toric architecture with a planar architecture.
Starting from $L=2$ and ending on a size $L$ code gives an overall overhead to emulating the torus of 6($\sum_{i=1}^{\log_2(L)-2} 2^{i+1}-2) = 6L-12\log_2L$, since from \Cref{fig:L2_to_L4} it can be seen that opposite edges need be made adjacent two times per iteration to enforce locality in it. Additionally, the time steps within each iteration must be implemented. Noticing that the red CNOT gates in \Cref{fig:L2_to_L4}(b) can be applied simultaneously with the gates in \Cref{fig:L2_to_L4}(a), this can be done in 6 time steps, leading to an additional $6\log_2(L) -6$ time steps in total in the RG encoder.

It is key to our routine to be able to ``spread out'' the qubits between each MERA step. We now show that this can be achieved in linear time by routing qubits through the planar grid. We firstly consider a single step of moving from a $2^k$ to a $2^{k+1}$ sized grid.

\begin{figure}[h]
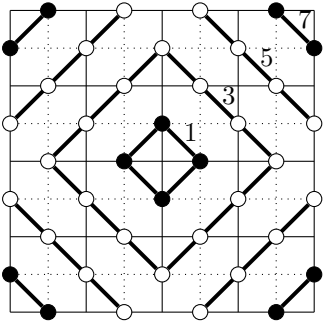

\begin{equation*}
\tikzfigscale{1}{figs/target}
\end{equation*}
\caption{$L=4$ code showing the circles of the $M_{i,j}$ metric. Qubits from the previous iteration are in black and new ones in white.}
\label{circles}
\end{figure}

Our first observation is that while the qubits lie on the edges of our $2^k \cross 2^k$ grid, one can subdivide this grid into one of dimensions $(2^{k+1}+1) \cross (2^{k+1}+1)$, such that the qubits lie on corners of this new grid, labelled by their positions $(i,j)$ with the centre of the grid identified with $(0,0)$. Under the taxicab metric we can measure the distance of qubits from the centre as $M_{i,j}:=|(i,j)| = |i|+|j|$ and one can check that qubits only ever lie at odd values of this metric, essentially forming a series of concentric circles with $M_{i,j} = 2n+1, n\in\mathbb{N}$. See \Cref{circles}.

A general routing step requires enlarging these circles such that the initial radii $R_I$ are mapped to final radii $R_F$ in the following fashion:
\begin{equation}\label{sorting matrix}
\begin{matrix}
R_I & & R_F & \hspace{1cm} STEPS \\
2^k-1 & \rightarrow & 2^{k+1}-1 & \hspace{1cm} 2^{k-1}\\
2^k-3 & \rightarrow & 2^{k+1}-7 & \hspace{1cm} 2^{k-1}-2\\
2^k-5 & \rightarrow & 2^{k+1}-9 & \hspace{1cm} 2^{k-1}-2 \\
2^k-7 & \rightarrow & 2^{k+1}-15 & \hspace{1cm} 2^{k-1}-4\\
& \vdots & &\hspace{1cm} \vdots\\
3 & \rightarrow & 7 & \hspace{1cm} 2\\
\end{matrix}
\end{equation}

Routing the qubits requires a series of SWAP gates to iteratively make the circles larger, e.g. $3\rightarrow 5\rightarrow 7$, the number of steps this requires is shown in table \ref{sorting matrix}. At the initial time step, it is only possible to move the outermost circle ($R_I=2^k-1$) since all smaller circles are adjacent. One can check though, that the number of steps required to move these smaller circles is sufficiently small that it is possible to start moving them at a later time step. We provide a framework for the required steps in equation \ref{routingroutine}.

Thus all the qubits can be moved in $2^{k-1}$ steps. Each step requires (possibly) simultaneous SWAP gates, each of which can be decomposed into three CNOT gates. Thus the overall run time of each iteration is $3\cdot 2^{k-1}$. To start from the $L=2$ base code and enlarge to a desired $L=2^m$ requires $\log_2(L)-1$ iterations and thus the overall run time for the routing routine is given by the geometric series $\sum_{k=1}^{\log_2(L)-1} 3\cdot 2^{k-1} = \frac{3}{2}(L-2)$.

Combining this with the time to emulate a toric architecture with a planar architecture, and the 10 time steps required to encode the $L=2$ base case using the Gottesman encoder (see Appendix~\ref{app:gottesman}), the total number of time steps required for the local RG encoder is $15L/2 - 6\log_2 L + 7 \sim O(L)$, where $L$ must be a power of 2.

\onecolumn
\begin{equation}\label{routingroutine}
\begin{matrix*}[r]
Time step & &&&&\\
1. &\hspace{0.95cm} 2^k-1\rightarrow 2^k+1 & WAIT & WAIT & \dots & WAIT \\
2. & 2^k+1\rightarrow 2^k+3 &\hspace{0.95cm} 2^k-3\rightarrow 2^k-1 & WAIT & \dots & WAIT\\
3. & 2^k+3\rightarrow 2^k+5 & 2^k-1\rightarrow 2^k+1 &\hspace{1.3cm} 2^k-5\rightarrow 2^k-3& \hspace{0.5cm}\dots & WAIT\\
\vdots & \vdots & \vdots & \vdots & \vdots & \\
2^{k-1}-1. & 2^{k+1}-5\rightarrow 2^{k+1}-3 & 2^{k+1}-9\rightarrow 2^{k+1}-7 & 2^{k+1}-13\rightarrow 2^{k+1}-11 & \dots & \hspace{0.5cm}3\rightarrow 5\\
2^{k-1}. & 2^{k+1}-3\rightarrow 2^{k+1}-1 & DONE & 2^{k+1}-11\rightarrow 2^{k+1}-9 & \dots & 5\rightarrow 7\\
\end{matrix*}
\end{equation}

\end{document}